\documentclass[11pt]{article}

\usepackage[table]{xcolor}
\usepackage{amsmath}
\usepackage{graphicx,psfrag,epsf}
\usepackage{enumerate}
\usepackage[numbers, sort&compress]{natbib}
\usepackage[hyphens]{url} 
\usepackage{ulem} 
\usepackage{standalone}
\usepackage{footnote}

\newcommand{\blind}{0}

\usepackage{makecell}
\usepackage{amsthm}
\usepackage{bm}
\usepackage{bbm}
\usepackage{mathtools}
\usepackage{collcell}
\usepackage{dsfont}
\usepackage{rotating}
\usepackage{wrapfig}
\usepackage[utf8]{inputenc}
\usepackage{tikz}
\usepackage{lettrine}
\usepackage{dblfloatfix}
\usepackage[english]{babel} 
\usepackage{amsfonts}
\usepackage{animate}
\usepackage{subfig}
\usepackage{floatflt}
\graphicspath{ {fig/} }
\usepackage[hyperfootnotes=false]{hyperref}

\usepackage[justification=centering]{caption} 
\usepackage{arydshln}
\usepackage{lipsum}
\usetikzlibrary{arrows,positioning,decorations.markings}
\usepackage{standalone}
\usepackage{tablefootnote}

\usepackage{etoolbox}
\makeatletter
\def\tikz@valign{c}
\tikzset{
  enforce alignment/.code={
    \csname if#1\endcsname
      \def\tikz@text@width
        {\pgfkeysvalueof{/pgf/minimum width}-2*(\pgfkeysvalueof{/pgf/inner xsep})}%
    \else
      \let\tikz@text@width\pgfutil@empty
    \fi},
  enforce alignment/.default=true,
  valign/.is choice,
  valign/top/.code=\def\tikz@valign{t},
  valign/center/.code=\def\tikz@valign{c},
  valign/bottom/.code=\def\tikz@valign{b},
  valign height/.initial=%
    \pgfkeysvalueof{/pgf/minimum height}-2*(\pgfkeysvalueof{/pgf/inner ysep})}
\patchcmd\tikz@fig@continue{\tikz@node@transformations}{%
  \pgfmathsetlength\pgf@x{\pgfkeysvalueof{/tikz/valign height}}%
  \pgf@y\ht\pgfnodeparttextbox
  \advance\pgf@y\dp\pgfnodeparttextbox
  \ifdim\pgf@y<\pgf@x
  \if\tikz@valign b%
    \advance\pgf@x-\dp\pgfnodeparttextbox
    \setbox\pgfnodeparttextbox\vbox to \pgf@x{\vfill\hbox{\unhbox\pgfnodeparttextbox}}%
  \else\if\tikz@valign t%
    \setbox\pgfnodeparttextbox\vbox to \pgf@x{\hbox{\unhbox\pgfnodeparttextbox}\vfill}%
  \fi\fi\fi
  \tikz@node@transformations}{}{}
\makeatother

\newcommand{\ov}\overline

\newcommand{\rig}\right
\newcommand{\lef}\left
\newcommand{\nf}\normalfont

\usepackage{forest} 
\usepackage{multirow}

\usepackage{subcaption} 

\usepackage{longtable}

\newcolumntype{C}[1]{>{\centering\arraybackslash}p{#1}}
\newcolumntype{R}[1]{>{\raggedleft\arraybackslash}p{#1}} 

\setcitestyle{authoryear,comma,aysep{,}}

\begin{document}

	\def\spacingset#1{\renewcommand{\baselinestretch}%
		{#1}\small\normalsize} \spacingset{1}

	
	\if0\blind
	{
		\title{\bf A data-driven merit order: Learning a fundamental electricity price model}
		\author{Paul Ghelasi\footnote{Corresponding author.} \footnote{Chair of Environmental Economics, esp. Economics of Renewable Energy. House of Energy Markets and Finance. University of Duisburg-Essen, Essen, Germany. Email addresses:  paul.ghelasi@uni-due.de (Paul Ghelasi), florian.ziel@uni-due.de (Florian Ziel).} , Florian Ziel\footnotemark[2] \\
		University of Duisburg-Essen, Germany}
		\maketitle
	} \fi

	\if1\blind
	{
		
		\bigskip
		
		\begin{center}
			{\LARGE\bf A data-driven merit order: Learning a fundamental electricity price model}
		\end{center}
	
	} \fi


	\begin{abstract}

		Electricity price forecasting approaches generally fall into two categories: data-driven models, which learn from historical patterns, or fundamental models, which simulate market mechanisms. We propose a novel and highly efficient data-driven merit order model that integrates both paradigms. 
		The model embeds the classical expert-based merit order as a nested special case, allowing all key parameters, such as plant efficiencies, bidding behavior, and available capacities, to be estimated directly from historical data, rather than assumed.
		We further enhance the model with critical embedded extensions such as hydro power, cross-border flows and corrections for underreported capacities, which considerably improve forecasting accuracy.
		Applied to the German day-ahead market, our model outperforms both classic fundamental and state-of-the-art machine learning models. It retains the interpretability of fundamental models, offering insights into marginal technologies, fuel switches, and dispatch patterns, elements which are typically inaccessible to black-box machine learning approaches.
		This transparency and high computational efficiency  make it a promising new direction for electricity price modeling.
	\end{abstract}
	
	\noindent%
	\vfill
	
	\spacingset{1.45} 

\newpage

\section{Introduction}


\subsection{Recent developments of electricity prices}
Accurate forecasting of electricity prices is essential for efficient operation and planning across the power system. Electricity producers rely on forecasts for power plant scheduling, buyers use them to optimize procurement strategies, and grid operators depend on them to maintain system stability and reduce the need for costly backup generation. 
\begin{wrapfigure}[15]{r}{8cm}
	\includegraphics[width=8cm]{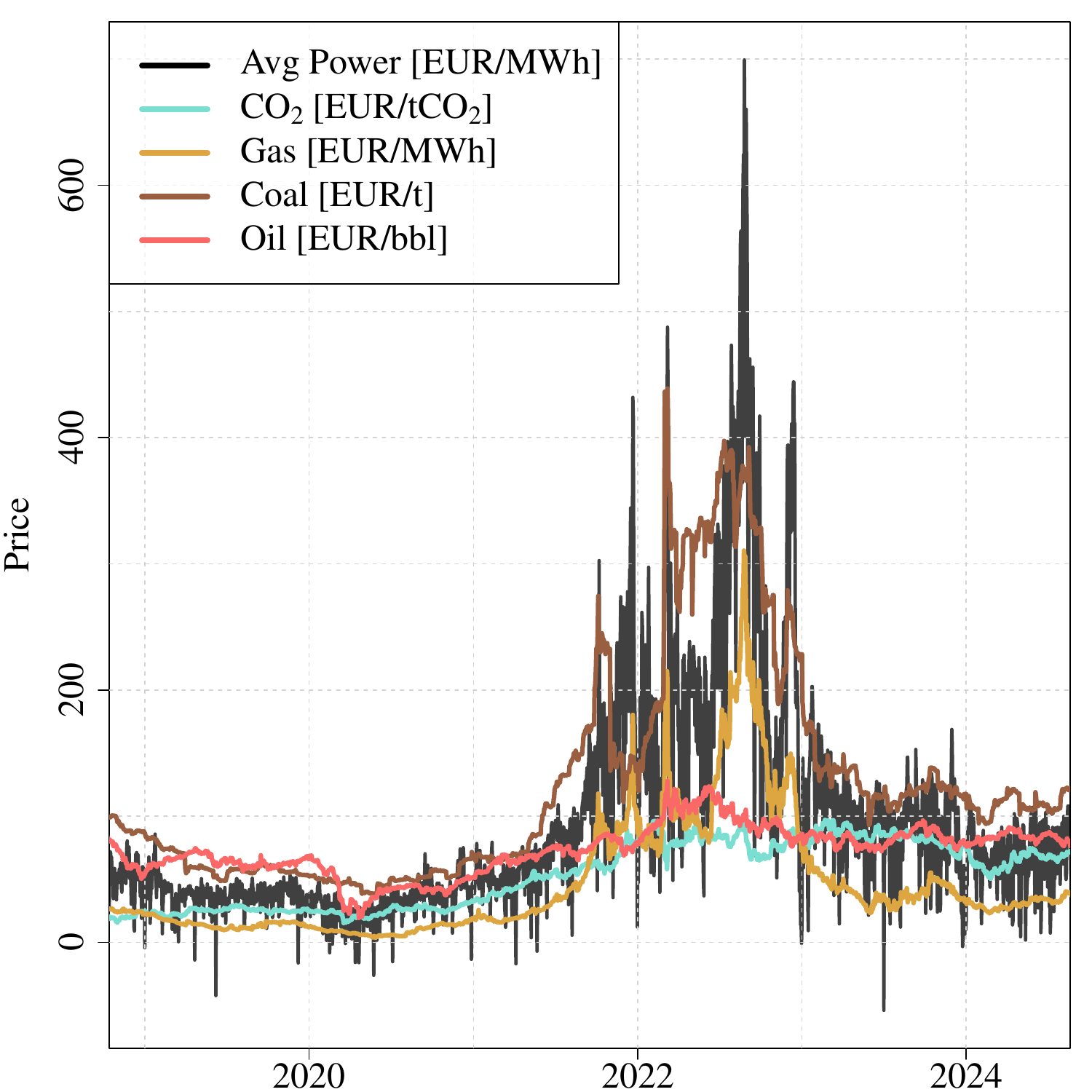}
	\caption{Fuel prices and daily average day-ahead power prices in Germany.}
	\label{fig:intro_price_fuels}
\end{wrapfigure}
As electricity markets evolve, the price time series has become increasingly volatile and complex, particularly with the growing share of renewables and the shifts following the European energy crisis.
In Europe, the most important electricity market is the day-ahead market, a blind daily auction that determines hourly prices for the following day. Each day at 12:00 CET, electricity prices are settled for every hour of the next day based on buy (bid) and sell (ask) volume-price offers submitted by wholesale market participants. Day-ahead prices are established through the aggregation of these bids, forming a market equilibrium where supply meets demand. This market serves not only as the primary platform for physical electricity trading but also as the underlying for the pricing of futures contracts and other derivatives. Its behavior reflects a broad range of physical and economic factors, including availability of conventional plants, renewable generation, demand fluctuations, fuel and $\text{CO}_2$ prices (see Figure \ref{fig:intro_price_fuels}).

\begin{wrapfigure}[15]{r}{8cm}
	\includegraphics[width=8cm]{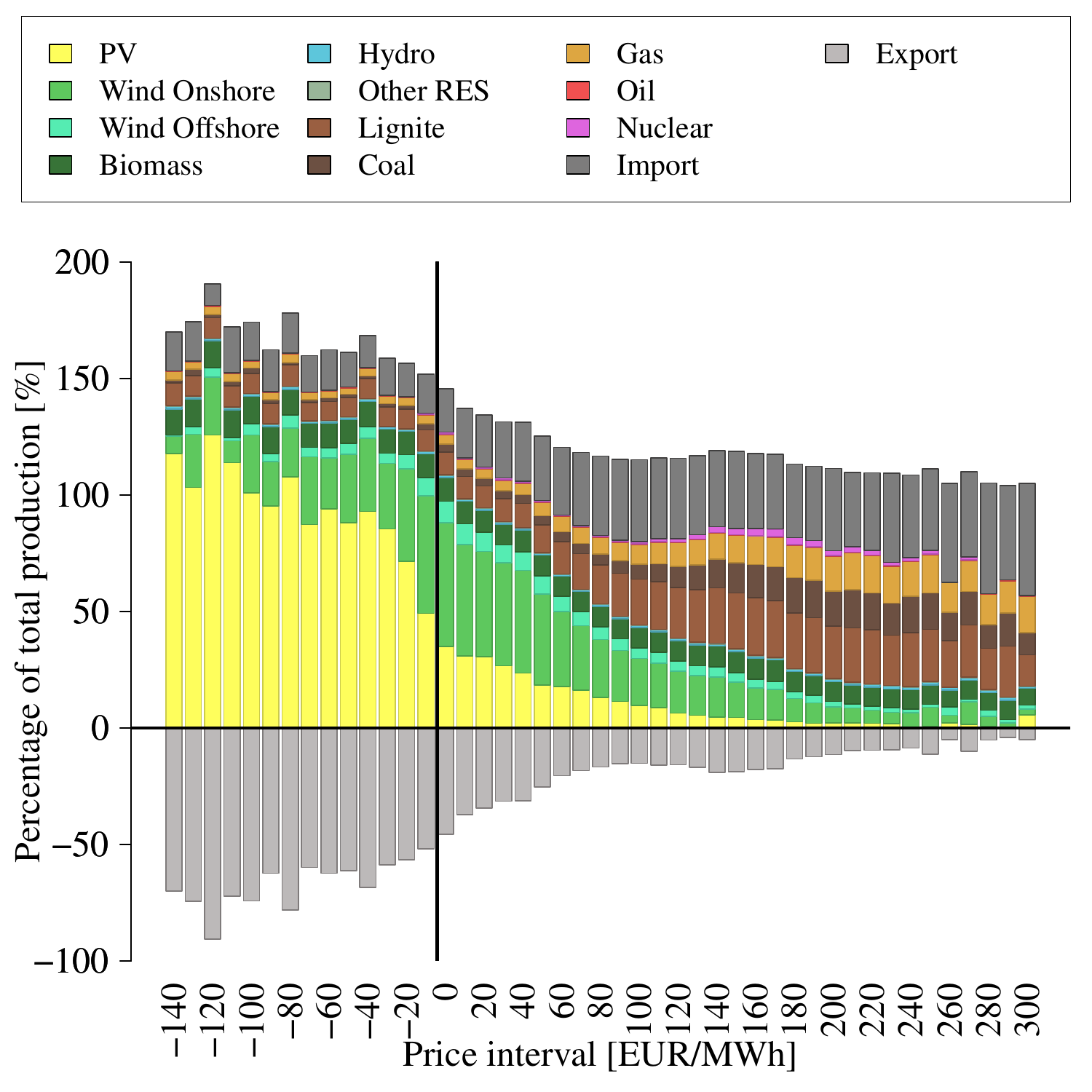}
	\caption{Average production given day-ahead power price intervals for 2023-2024 in Germany.}
	\label{fig:intro_price_intervals}
\end{wrapfigure}

\begin{figure}[hbtp]
\subfloat[Occurrences of sequential negative prices.]{
	\includegraphics[width=0.49\textwidth]{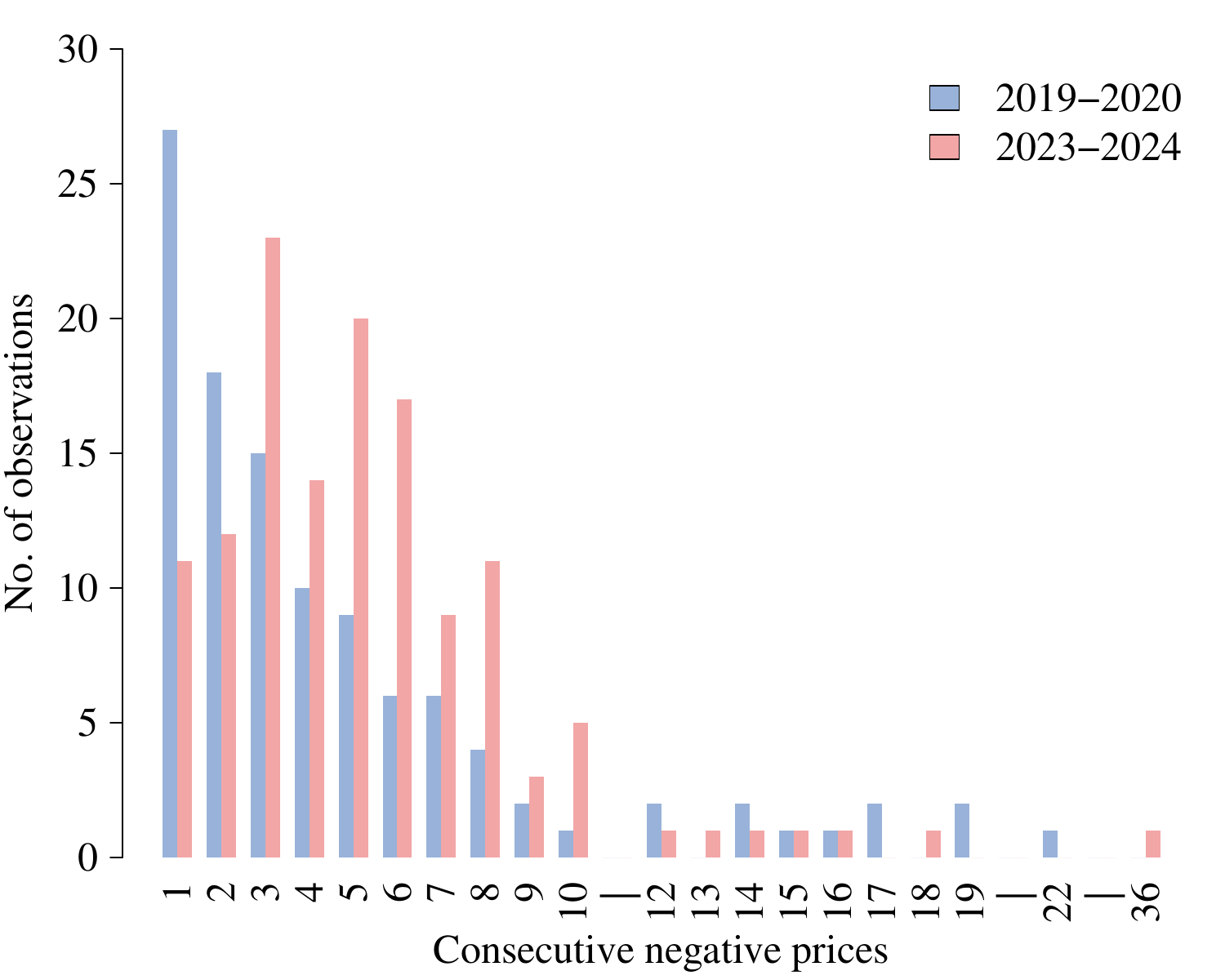}}
\subfloat[Histograms of prices for selected periods.]{
	\includegraphics[width=0.49\textwidth]{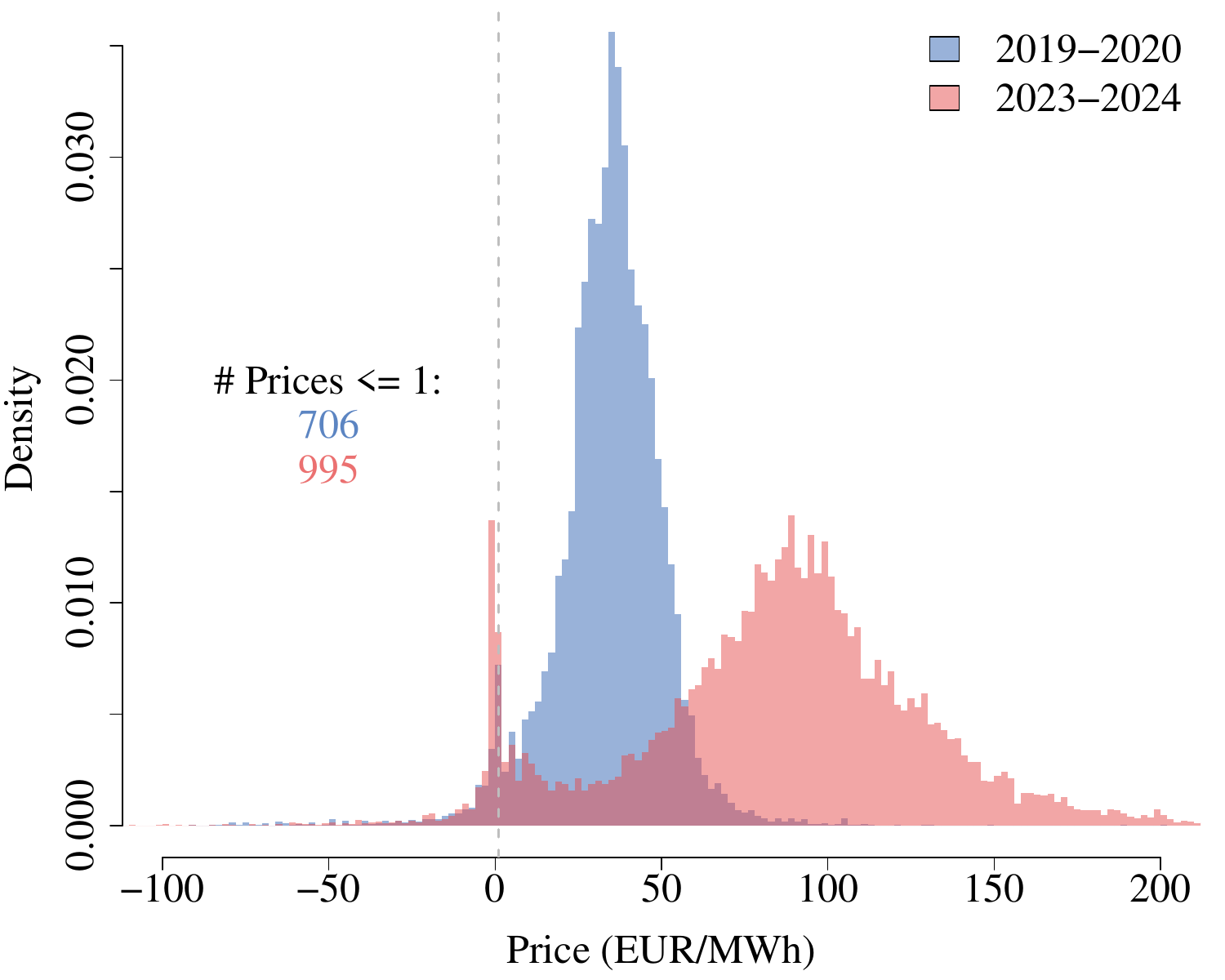}}
	\caption{Evolution of day-ahead electricity prices in Germany}
\label{fig:intro_negative_prices}
\end{figure}

The integration of large-scale intermittent renewable energy, such as wind and solar, has contributed to more frequent occurrences of very low or even negative prices. This trend is particularly evident in countries like Germany, where over 50\% of total installed capacity are renewable energy sources (RES). This is mainly because many RES operate under subsidy schemes that allow them to bid at negative prices. Even without subsidies, their marginal costs are close to zero, enabling them to bid at very low or zero prices. As a result, RES dominate electricity generation during low-price periods, while conventional power plants tend to ramp up when prices are higher (see Figure \ref{fig:intro_price_intervals}).
These conditions have contributed not only to a rise in negative prices during periods of high RES generation but also to an increase in the number of consecutive hours with negative prices (see Figure \ref{fig:intro_negative_prices}). Another important development is the increasing concentration of prices around zero, along with the tendency for such low prices to occur in consecutive price clusters. Figure \ref{fig:intro_negative_prices}.b shows that, following the European energy crisis, the electricity price distribution remained bimodal, but the gap between the two modes widened. This was driven by both an upward shift in the average price and a higher frequency of low or negative price occurrences. The two modes reflect different pricing regimes: one associated with high prices, typically linked to fossil fuel-based generation, and another with low or negative prices, typically observed during periods of high renewable output.

These dynamics call for price models that can address two key requirements: first, the ability to capture regime shifts in market behavior; and second, the flexibility to adapt to changing system conditions, such as the increasing clustering of prices. This implies a need for models that go beyond statistical pattern recognition by incorporating fundamental drivers of electricity prices.

\subsection{Electricity price forecasting models}

\subsubsection{Data-driven vs. fundamental models}
\begin{figure}[htb]
    \centering
	\includegraphics[width=1\textwidth]{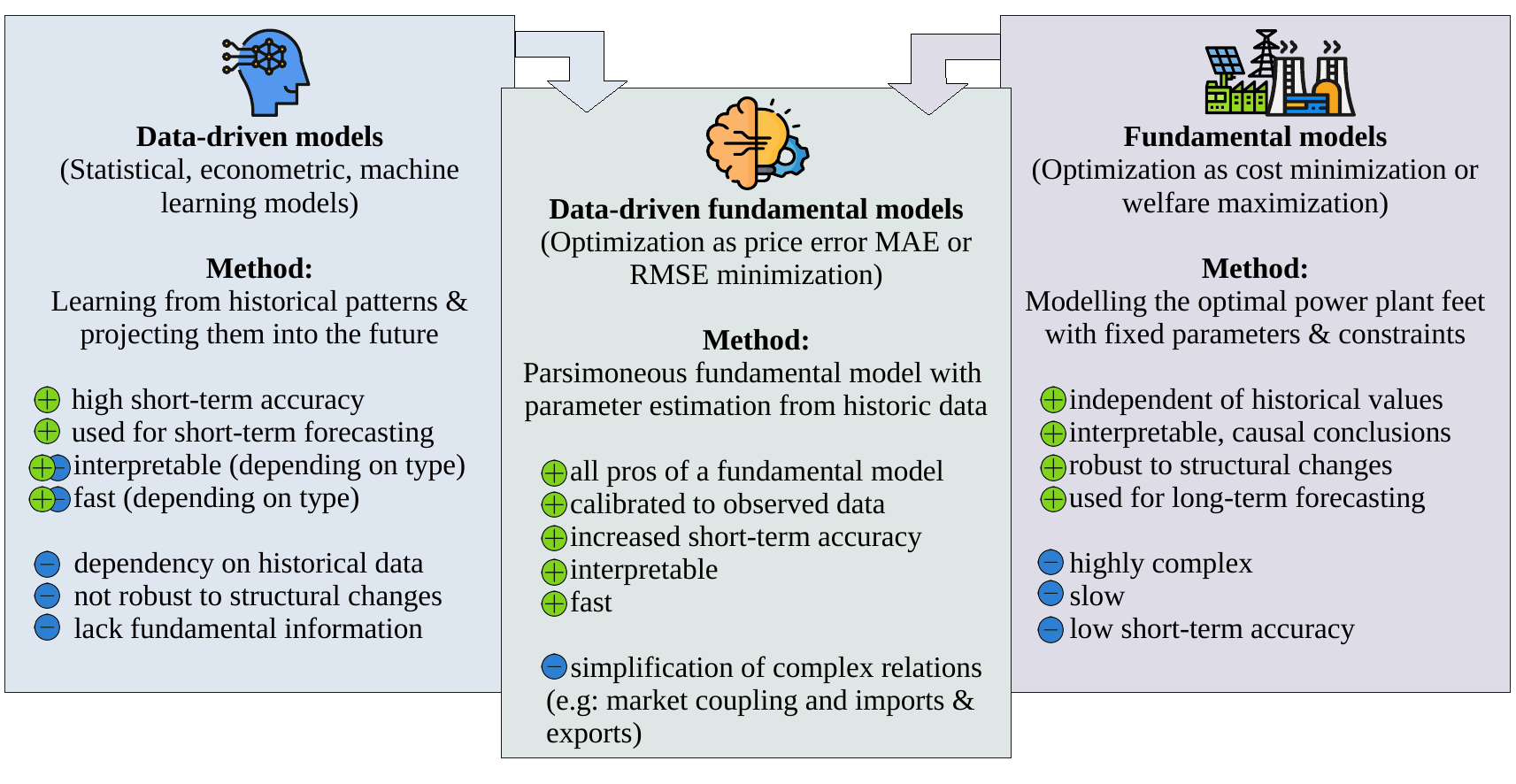}
	\caption{Comparison between existing model classes, and the newly introduced data-driven fundamental model type as a hybrid combination of the two.}
	\label{fig:sketch}
\end{figure}

Electricity price forecasting methods can be broadly classified into two categories: \emph{fundamental} and \emph{data-driven models} \citep{petropoulos2022forecasting, weron2014electricity, ziel2016electricity}. Each have their strengths and weaknesses which are summarized in Figure \ref{fig:sketch}.
\emph{Data-driven} models include classical econometric, statistical, as well as advanced machine learning approaches. These models use historical data to forecast power prices by identifying past patterns and projecting them into the future \citep{hastie2009elements, lutkepohl2005new, brockwell2002introduction}. These are primarily used for short-term forecasting \citep{cuaresma2004forecasting, narajewski2020econometric, uniejewski2019understanding, petropoulos2022forecasting} because they effectively capture the high volatility of power prices and run efficiently, especially with linear models and online learning algorithms \citep{wintenberger2017optimal, cesa2021online, adjakossa2024kalman, ziel2022smoothed}.

\emph{Fundamental} models on the other hand simulate the electricity market mechanism and power plant fleet within a geographic area \citep{ringkjob2018review, brown2017pypsa, petropoulos2022forecasting, beran2021multi}. These are also referred to as or contain structural models, market-based equilibrium models, cost optimization models and agent-based models.
These models are mostly optimized by minimizing overall power production costs or maximizing overall welfare, often for complex networks spanning multiple countries.
Their outputs range from optimal power plant schedules to capacity buildouts and power plant mixes, with the price of power typically being only one of the outputs, calculated for example as the shadow price or dual variable of the demand constraint \citep{marcos2019electricity, brown2017pypsa}. Traditional fundamental models require detailed technical information and complex operational constraints, making them computationally expensive and unsuitable for short-term forecasts that need frequent updates. Moreover, for short-term forecasting, they are often described as too "flat" meaning that they fail to capture the high volatility of power prices \citep{pape2016fundamentals, marcos2019electricity, beran2021multi} (see Figure \ref{fig:intro_forecast}). However, unlike data-driven models they, are able to capture price clusters, which often form around zero in markets with a high share of RES (see Figure \ref{fig:intro_forecast}). Examples of open-source fundamental and agent-based models include \textit{PyPSA} \citep{brown2017pypsa}, \textit{Calliope} \citep{pfenninger2018calliope}, \textit{OSeMOSYS} \citep{howells2011osemosys}, \textit{AMIRIS} \citep{schimeczek2023amiris} and \textit{ASSUME} \citep{harder2025assume}.

\subsubsection{Limitations of pure data-driven and fundamental models}
While data-driven models excel in short-term forecasting accuracy, they are highly sensitive to the historical data used for training and can fail under regime changes, even if the market mechanism remains unchanged. This issue is particularly problematic for long-term forecasts \citep{ghelasi2025day}. 

\begin{wrapfigure}[29]{r}{8cm}
	\includegraphics[width=8cm]{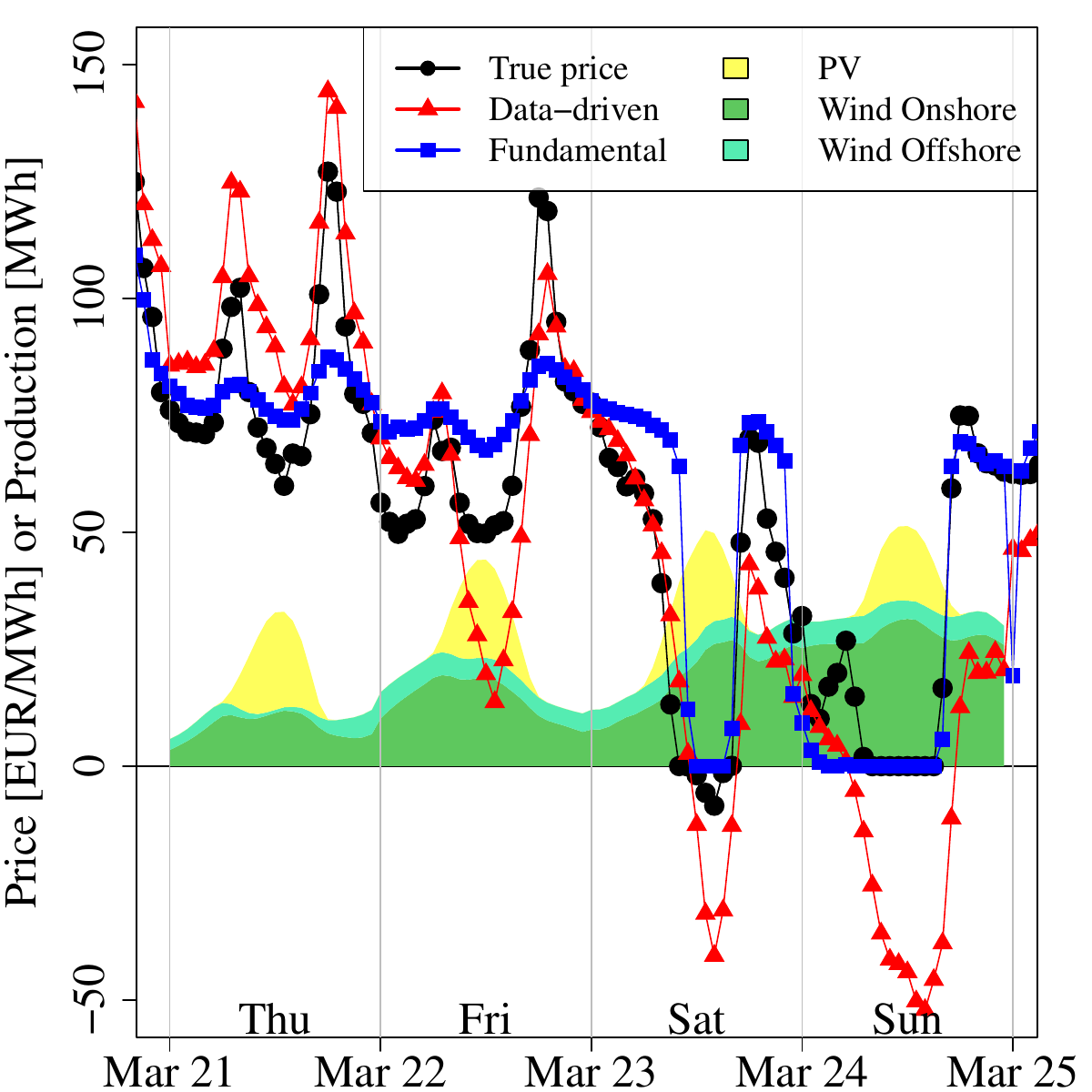}
	\caption{Fundamental vs data-driven models forecasts for Germany (2024).}
	\label{fig:intro_forecast}
		\includegraphics[width=8cm]{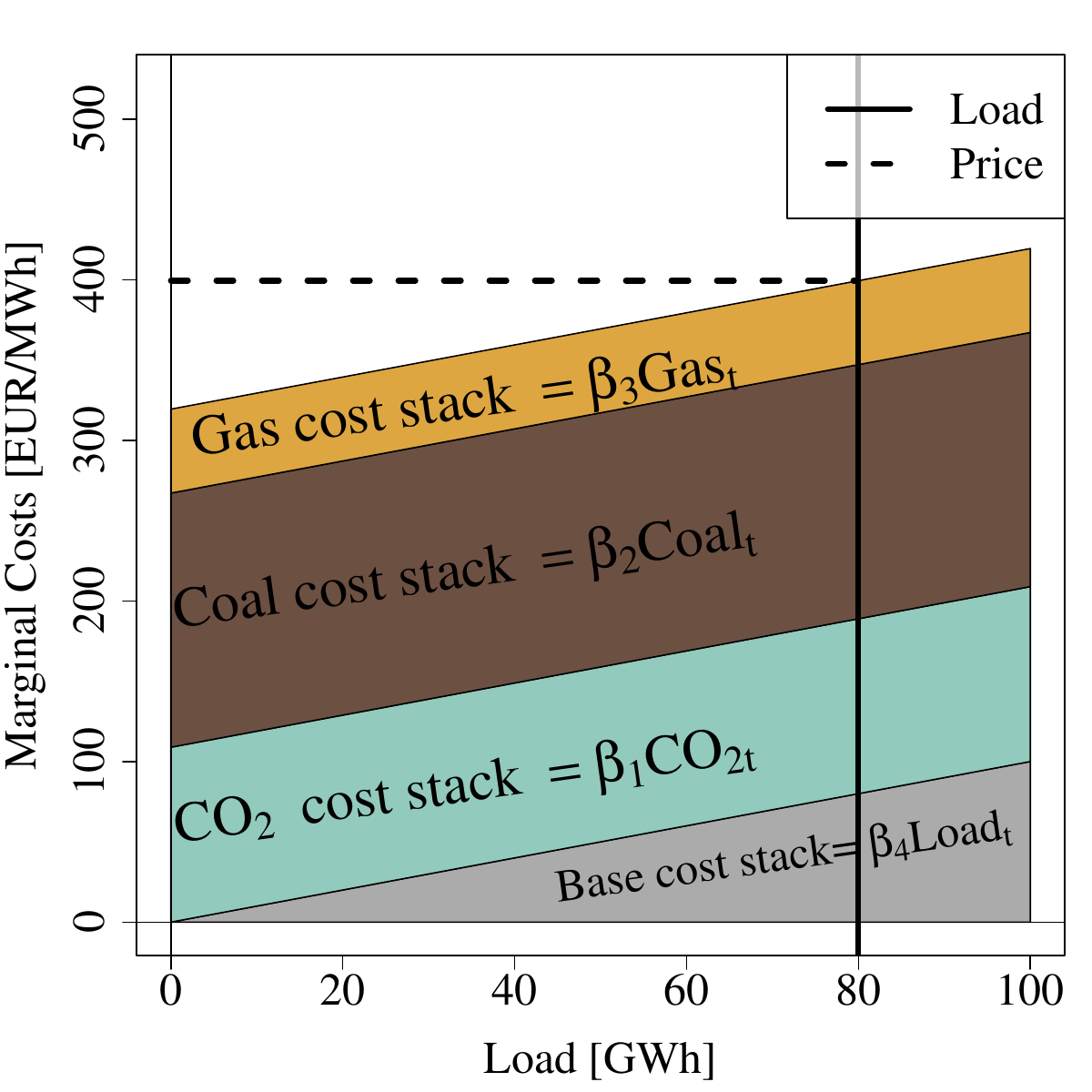}
	\caption{Merit order representation of: \\ \small{$\textrm{Price}_t = {\beta}_1 \textrm{CO}_{2,t} + {\beta}_2 \textrm{Coal}_t + {\beta}_3 \textrm{Gas}_t + {\beta}_4 \textrm{Load}_t + \varepsilon_t$}}
	\label{fig:intro_mo_expert}
\end{wrapfigure}

For instance, consider a market consisting only of gas and coal power plants, with a simple linear econometric model predicting prices based on fuel costs, as illustrated in Figure \ref{fig:intro_mo_expert}.
The figure shows the model's merit-order (MO) representation, created by rearranging and plotting the estimated coefficients, demonstrating that any econometric price model can be visualized in this form. 
When a new technology such as wind power is introduced, the data-driven model cannot capture its effect, and therefore cannot predict prices accurately if the impact is substantial, due to the lack of historical data needed to estimate a wind-related coefficient. 
Although pure data-driven models can very well detect the pattern of lower prices during periods of high renewable energy (RES) generation, they lack a fundamental explanation for this effect, namely, that RES has near-zero marginal costs. As a result, they can fail to capture the clustering of prices around zero and tend to overshoot actual prices, as seen in Figure \ref{fig:intro_forecast}. This is expected, as such models learn only from historical patterns and lack a fundamental understanding of the underlying market mechanisms.

In contrast, a fundamental model does not rely on historical data. It simulates the market mechanism and can generate prices based on current or expected inputs such as forecasted RES expansion or fuel prices, provided the underlying mechanism remains unchanged. If a new technology like wind is introduced, its effect can be easily modeled by specifying its current cost parameters and available capacity (see the merit order model in the sections below). However, fundamental models are often too slow or inflexible for short-term forecasting. This highlights the need for a hybrid approach that combines the strengths of both methods.

\subsubsection{Combined models in literature}
There have been some attempts in literature to combine the strength of both modelling approaches. There are two main categories, first the \textit{hybrid models} and \textit{inverse optimizations}. Table \ref{tab:papers1} summarizes some of the most relevant papers in these categories by their methodology, data, period and performance. The list was compiled and manually filtered to only include papers that explicitly also model electricity prices.
\citep{bello2016medium, beran2021multi, gonzalez2011forecasting, marcos2019electricity, gabrielli2022data}.
\begin{table}[htpb]
	\setlength{\tabcolsep}{1pt}
	\centering
	\scalebox{0.6}{
		\setlength\extrarowheight{-3pt}	
		\begin{tabular}{|>{\centering\arraybackslash}p{3cm}|>{\centering\arraybackslash}p{2cm}|>{\centering\arraybackslash}p{4cm}|>{\centering\arraybackslash}p{4cm}|>{\centering\arraybackslash}p{2cm}|>{\centering\arraybackslash}p{5cm}|>{\centering\arraybackslash}p{5cm}|}
			\hline
			\textbf{Paper} & \textbf{Market} & \textbf{Methods} & \textbf{Data} & \textbf{Period} & \textbf{Measure} & \textbf{Scope} \\
			\hline
			\citeauthor{gabrielli2022data}, \citeyear{gabrielli2022data} , \citep{gabrielli2022data} 			& UK 				& Hybrid: Market model \& Linear regression \& Gaussian process regression \& Neural networks 	& Autoregressive, seasonalities, demand, generation, RES, fuel prices & 2015 - 2019 & MAPE & Forecasting of prices over 10 years ahead \\
			\hline
			\citeauthor{beran2021multi}, \citeyear{beran2021multi}, \citep{beran2021multi} 						& Germany 			& Hybrid: Parsimonious fundamental model \& ARX & Autoregressive, seasonalities, load, RES,  	& 2014 - 2016 & MAE, WMAE & Forecasting of prices up to 1 week ahead \\
			\hline
			\citeauthor{marcos2019electricity}, \citeyear{marcos2019electricity}, \citep{marcos2019electricity} & Iberian peninsula & Hybrid: Cost-production optimization model \& Neural network & Autoregressive, demand, RES, seasonalities, power plant data, transmission constraints & 2017 &  MAPE, MAE, RMSE & Forecasting of prices day and week-ahead  \\
			\hline
			\citeauthor{bello2016medium}, \citeyear{bello2016medium}, \citep{bello2016medium} 					& Spain 			& Hybrid: Market equilibrium model \& Quantile regression & Autoregressive, fuel prices, demand & 2012 - 2014 & MAPE & Forecasting of prices up to 1 year ahead \\
			\hline
			\citeauthor{gonzalez2011forecasting}, \citeyear{gonzalez2011forecasting}, \citep{gonzalez2011forecasting} 		& UK 	& Hybrid: Supply stack model \& Logistic smooth transition regression model & Autoregressive, fuel prices, demand & 2008 & MAPE & Forecsating of day-ahead prices \\
			\hline

			\citeauthor{liang2023data}, \citeyear{liang2023data}, \citep{liang2023data} 		& NYISO (US) 	& Inverse optimization of a cost minimization problem & Generation, transmission lines, prices & 2018 & Error between recovered prices and observed market clearing data using gradient decent & Estimation parameters: marginal offer price per generator \\
			\hline
			\citeauthor{birge2017inverse}, \citeyear{birge2017inverse}, \citep{birge2017inverse} 		& MISO (US) 	& Inverse optimization of an economic dispatch problem & Demand, capacities, transmission data, prices & 2012 & Calibration to location marginal prices (LMPs) using custom algorithm & Estimation of parameters: transmission utilization coefficients, locational loss factors, transmission line capacities \\
			\hline
			\citeauthor{chen2017strategic}, \citeyear{chen2017strategic}, \citep{chen2017strategic} 		& Simulation	& Inverse optimization of a profit maximization problem & Production costs  & - & Minimization of the sum of duality gaps  & Extracting optimal bidding behaviors from the day-ahead electricity market. Estimation of parameters: cost function of suppliers \\
			\hline
			\citeauthor{ruiz2013revealing}, \citeyear{ruiz2013revealing}, \citep{ruiz2013revealing} 		& Simulation 	& Inverse optimization of a cost minimization problem & Marginal costs, generation, transmission capacities, ramp up/down limits & - & Discrepancy between observed market outcome and predicted outcomes & Estimated parameters: offer prices of rival producers \\
			\hline
		\end{tabular}}
	\caption{Papers on electricity price modelling where data-driven and fundamental models are used in a combined way.}
	\label{tab:papers1}
\end{table}

\textbf{Regime-switching models.} In the past, \textit{Markov regime-switching} models and variations thereon have been proposed to capture the price-switching behavior of electricity markets, and the literature on them is extensive. These models initially focused on modeling price spikes rather than low or negative prices, reflecting the conditions of the early 2000s when renewable energy penetration was limited and peak prices were the primary concern \citep{weron2004modeling, haldrup2006regime, mount2006predicting, karakatsani2008intra}, but more timely applications have been tried as well \citep{kapoor2023analyzing}. However, they are strictly data-driven and lack fundamental interpretability, offering limited insight into the underlying market mechanisms. Fundamental models, while not regime-switching models, can still capture this type of behavior intrinsically as they simulate the underlying mechanism.

\textbf{Hybrid models.}
In the electricity price forecasting literature, the term \textit{hybrid} is often used to describe approaches that combine different model types. A particularly relevant subset involves the combination of fundamental models with data-driven methods. In such cases, the hybridization is nevertheless implemented as a post-processing step, meaning that the output of the fundamental model is used as an additional input to the data-driven model. This form of integration is \textit{extrinsically} hybrid, as the two components remain functionally and structurally separate. Table \ref{tab:papers1} summarizes the papers in this category.

For example, \citet{beran2021multi} use forecasts from a parsimonious fundamental model as inputs to a set of expert data-driven models, specifically ARX models, following the framework of \citet{ziel2018day}. This improves short-term price forecasting in the German market. Similarly, \citet{gonzalez2011forecasting} apply a merit-order-based fundamental model to generate price forecasts for the UK market, which are then fed into ARMA and LSTR (logistic smooth transition regression) models. Comparable procedures are used by \citet{marcos2019electricity} and \citet{gabrielli2022data} for the Spanish and UK markets, respectively. Although these methods have shown improvements in forecast accuracy, the coupling between the fundamental and data-driven models is limited to a feed-forward mechanism, and the integration is not realized at a structural or algorithmic level.
Other numerous approaches also labeled as hybrid \citep{zhang2020adaptive, alkawaz2022day, yang2019hybrid, shafie2011price, wan2013hybrid, huang2021novel} involve combinations of only data-driven methods, such as neural networks, support vector machines, or ensemble learning techniques, without including any market-based or physical modeling. While these models can effectively capture statistical regularities in the data, they may lack interpretability and may struggle to generalize in the presence of structural market shifts or regime changes.

The major limitation of this approach is that the data-driven model functions purely as a post-processing step, lacking any integration of fundamental constraints or causal interpretability. As a result, it offers no guarantees of generalization to new market conditions or of capturing structural effects such as price clustering driven by RES.

\textbf{Inverse optimization.}
A more intrinsic approach to combining the two modeling types is offered by \textit{inverse optimization} methods. The term \textit{inverse} can refer to multiple approaches, as summarized by \citep{esser2025multi}: \textit{inverse problem theory} refers to the estimation of a cause given an effect, instead of the other way around \citep{tarantola2005inverse, kirsch2011introduction}. This includes \textit{parameter estimation} or \textit{inverse modelling} \citep{da2023computational, groetsch1993inverse, richter2015inverse}, where the unknown cause is inferred given a known effect and model, as well as \textit{function estimation}, which refers to the estimation of the model itself given a known cause and effect \citep{gonzalez2022building}. \textit{Inverse simulation} refers to using simulations to replicate observed behavior \citep{kurahashi2018model, murray2000inverse}. Newly \textit{inverse optimization} is also used refer to problems where variables traditionally treated as inputs are treated as outputs instead and optimized for \citep{esser2025multi}.

In the context of energy systems modeling, inverse optimization typically refers to recovering market data parameters from observed market outcomes \citep{birge2017inverse}. This involves finding optimal model parameters that best reproduce observed outcomes—essentially solving the inverse of the classical forward problem, where fixed inputs produce modelled outputs. In inverse optimization, inputs (i.e., parameters) are adjusted to minimize the error between the model's output and actual market data. This aligns with the essence of data-driven models, where parameters are estimated by minimizing output error, rather than being fixed or based on expert assumptions.

In the context of energy systems modeling it is sufficient to note that \textit{inverse optimization} refers to recovering market data parameters from an actual market outcome \citep{birge2017inverse}. This means finding the optimal fundamental model parameters given the observed data. This is in contrast to the classical forward problem, as discussed for the \textit{hybrid models}, where the fundamental model has fixed inputs and produces modelled outputs. Inverse optimization methods reverse this process by finding those input values, in our case parameters, that minimize the error between the output of the model and the true value. In inverse optimization, inputs, i.e., parameters, are adjusted to minimize the error between the model's output and actual market data. This aligns with the essence of data-driven models, where parameters are estimated by minimizing output error, rather than being fixed or based on expert assumptions.

Literature on inverse optimization in electricity markets remains limited but spans various applications, such as recovering consumption decision models \citep{kovacs2021inverse}, estimating unobserved demand components \citep{esteban2024estimating}, and determining agents' objective functions \citep{nguyen2020cooperative}. A significant portion focuses on reverse-engineering optimal bidding strategies in electricity markets \citep{saez2016data, saez2017short, gallego2017inverse, bertsimas2015data, fernandez2021inverse, chen2017strategic, chen2019learning}, particularly on the demand side. Most of these works aim to estimate the bidding behavior or optimal operation of price-responsive loads based on market outcomes, though many rely on simulations rather than real-world data.

Some studies have applied inverse optimization to price modeling or forecasting, as summarized in Table \ref{tab:papers1}. \citet{liang2023data} applied inverse optimization to a cost-minimization model of the NYISO market to recover marginal offer prices at the generator level, minimizing the error between modelled and observed prices. \citet{chen2017strategic} used a simulation-based approach to estimate optimal bidding behavior in the day-ahead market. \citet{birge2017inverse} inferred technical parameters, such as utilizations and transmission line capacities, by minimizing errors between modeled and actual prices. Similarly, \citet{ruiz2013revealing} recovered unobserved offer prices of rival producers using a cost-minimization model.

However, none of these studies apply inverse optimization to a fundamental model with the sole purpose of real-world electricity price forecasting, in a data-driven forecasting framework. Most rely on simulations and involve complex optimization frameworks with numerous constraints, limiting their suitability for short-term forecasting due to computational demands.

\subsection{Novelties}

The literature review above show that while regime switching models, hybrid models have been proposed in the context of electricity price forecasting, they do not fully leverage the strengths of both fundamental and data-driven approaches. Inverse optimization methods, while promising, have not been applied for real-world electricity price forecasting in a data-driven framework. We fill this gap by proposing a novel, highly-efficient \emph{data-driven fundamental} \textbf{merit order model} tailored to electricity price forecasting, which is in essence an inverse optimization approach applied to a fundamental model, however, we essentially use it exactly as a data-driven econometric or machine learning model is used, i.e., estimating the parameters by minimizing the forecasting error on the training set and then using the model to forecast prices on a test set. This approach allows us to retain the causal interpretability of fundamental models and their implicit capability of capturing market effects such as regime switches and RES expansion, while achieving the accuracy and computational efficiency of data-driven models.

This approach intrinsically combines the strengths of both fundamental and data-driven models, as summarized in Figure \ref{fig:sketch}. The key novelties of this approach are:

\begin{enumerate}
	\setcounter{enumi}{0}
	\item \textbf{Data-driven fundamental model:} We develop a highly efficient fundamental merit order model that follows the structure of data-driven models by directly optimizing parameters to minimize prediction errors. This allows the model to retain the causal interpretability of fundamental models while achieving the accuracy and flexibility typical of data-driven approaches.
	\item \textbf{Forecasting-focused objective:} The model is specifically designed for electricity price forecasting by minimizing a dedicated prediction error metric (e.g. MAE) in the loss function, unlike traditional fundamental models where price is one of many outputs.
	\item \textbf{Market-embedded systems:} The model nests the classical fundamental structure by initializing parameters based on expert knowledge. These initial values serve as a starting point and can be further optimized, allowing for data-based refinement while preserving a link to established domain knowledge. In this way, if the optimal solution is the classical model, the optimization algorithm will converge to it.
\end{enumerate}

We apply this model to day-ahead electricity price forecasting in Germany and demonstrate that it outperforms both classical fundamental and data-driven models of similar complexity. To our knowledge, this is the first implementation of such an approach.

This paper is structured as follows.
Section \ref{sec:data} introduces the data and insights into power price formation, motivating the use of a fundamental model as the starting point.
Section \ref{sec:model} introduces the fundamental \emph{merit order} (MO) or \emph{supply stack} model and the formulation of the data-driven approach.
Section \ref{sec:result} presents parameter estimation approaches for the MO model and compares the results to state-of-the-art benchmarks which are discussed in Section \ref{sec:discussion}.
Finally, section \ref{sec:conclusion} concludes and outlines future research directions.

\section{Data}
\label{sec:data}

\subsection{Capacities, generation, exports, imports, fuel prices}
\begin{figure}[htbp]
    \centering
	\includegraphics[width=1\textwidth]{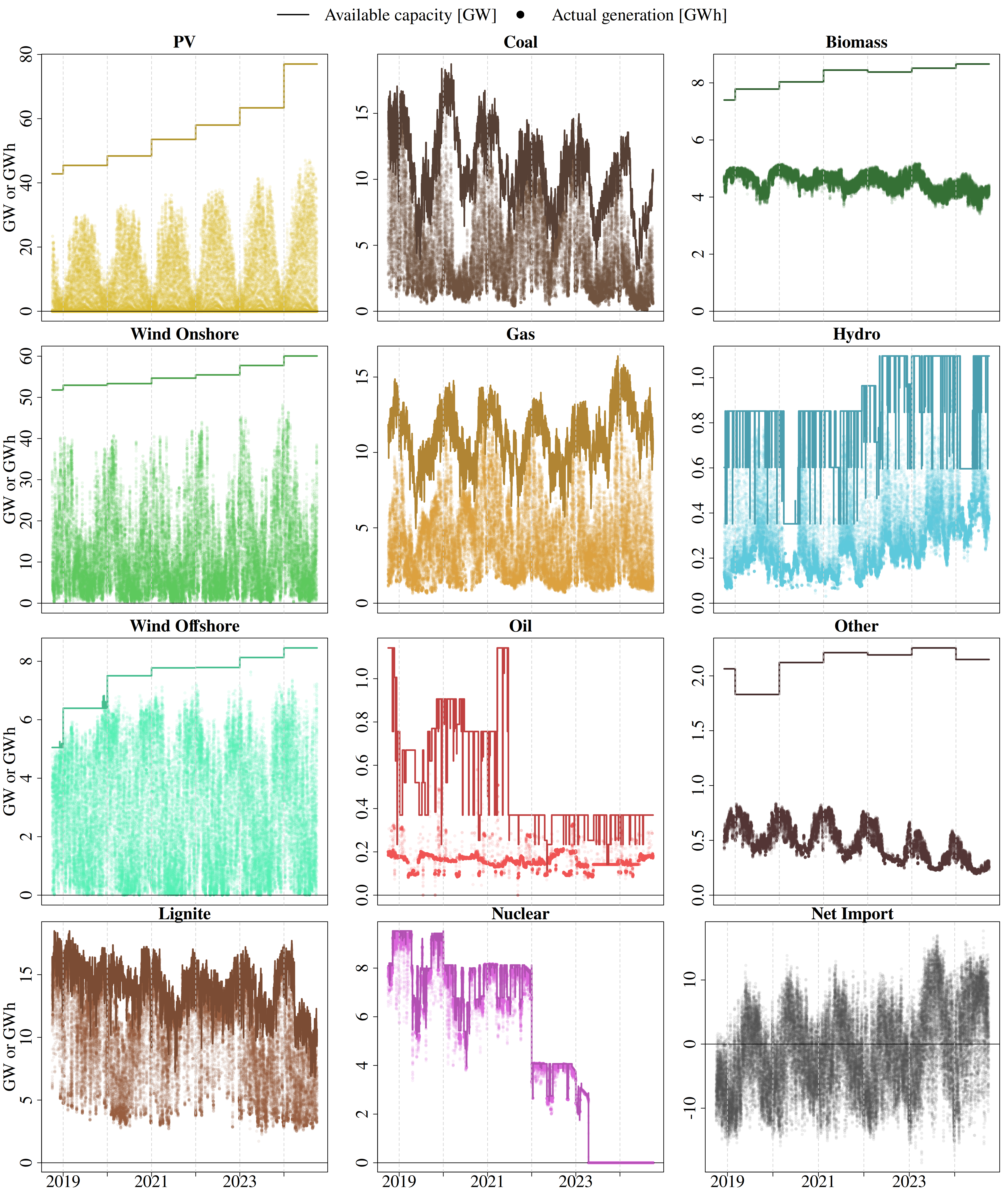} 
	\caption{Generation and available capacities by power plant type in Germany aggregated from unit-level reported data. May contain underreported data. }
	\label{fig:data_cap_gen}
\end{figure}

In this study, we use data from Germany, Europe's largest electricity market with a balanced energy mix across major sources. Unlike countries like France, dominated by nuclear power, Germany's diverse generation allows for meaningful analysis of both renewables and conventional technologies, making it ideal for applying the merit order model.
A key component of the merit order model is the available capacities and generation of power plants. This data is sourced from the \textit{ENTSOE Transparency} platform, which reports unavailabilities, installed capacities, and hourly generation at the power plant unit level. We aggregated this data by power plant type to produce the values shown in Figure \ref{fig:data_cap_gen}.
Net imports are calculated as the difference between Germany's total imports and exports, which have increased significantly since the nuclear phase-out on April 16, 2023. Available capacities show some volatility due to planned maintenance and temporary unplanned outages. Yearly patterns are evident in conventional plants like coal and gas, where maintenance occurs in summer when prices are typically lower due to high photovoltaic (PV) infeed.
Day-ahead forecasts for RES production and load are also provided on the platform. However, since day-ahead forecasts for net import and hydro are unavailable, we generate our own forecasts for these categories.

For convenience, we group all power plant types in a set $\mathcal{T}$, with subsets for RES plants $\mathcal{R}$ and conventional plants $\mathcal{C}$:
\begin{align}
	\mathcal{T}  = & \{ \text{Gas}, \text{Coal}, \text{Lignite}, \text{Oil}, \text{Nuclear}, \text{PV}, \text{Wind Onshore}, \text{Wind Offshore}, \text{Biomass},  \nonumber \\ 
	& \text{Other RES}, \text{Other}\},  \\
	\mathcal{R} = & \{ \text{PV}, \text{Wind Onshore}, \text{Wind Offshore}, \text{Biomass}, \text{Other RES} \}, \nonumber \\
	\mathcal{C} = & \mathcal{T} \setminus \mathcal{R}. \nonumber
\end{align}
Another essential input is fuel prices, including gas, coal, and $\textrm{CO}_2$ European Emission Allowances (EUA). These prices are sourced as front-month futures from the \textit{Refinitiv EIKON} information platform (see Figure \ref{fig:intro_price_fuels}).

\subsection{Power plant parameters}

Power plant parameters such as the efficiency, $\textrm{CO}_2$ intensity factors and marginal costs are further critical inputs for the merit order model. These parameters are traditionally estimated by experts and treated as fixed in the model. In the next section, we introduce these parameters along with the model, as they will serve as starting values for our estimation procedures (see Table \ref{tab:eff_co2}).

\subsection{Marginal costs}

Determining which power plant type sets the price at any given time is challenging as it requires analyzing the generation stack relative to the day-ahead price.
Figure \ref{fig:intro_price_intervals} illustrates the average production percentages for different price intervals in 2023–2024. Conventional plants like gas, coal, and lignite ramp up production during higher price periods, while RES producers bid even at negative prices due to subsidy schemes. Some conventional plants also offer power at negative prices due to high ramp-down costs, contractual obligations, or combined heat and power (CHP) mechanisms.
Imports increase during high-price periods, while excess RES production is exported during low-price periods. Certain plant types, such as gas, lignite, coal, oil, nuclear, and hydro, exhibit a clear must-run stack, as their total generation never drops to zero (see Figures \ref{fig:intro_price_intervals} and \ref{fig:data_cap_gen}).

\section{Model}
\label{sec:model}
\subsection{Merit order model}
\label{sec:mo_model}
\begin{figure}[htb]
    \centering
	\includegraphics[width=0.8\textwidth]{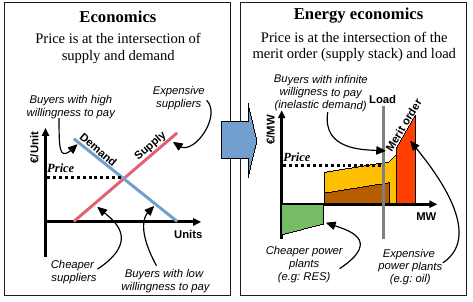}
	\caption{Economic theory behind the merit order model.}
	\label{fig:mo_sketch_economics}
\end{figure}

The \textbf{merit order model} is a basic fundamental approach to power price modeling, based on standard economic theory applied to the energy market. Figure \ref{fig:mo_sketch_economics} shows this relationship. 
Also referred to as the \textbf{supply stack model}, it simulates the electricity market by constructing the supply curve as the marginal costs of power plants sorted from lowest to highest. This is the equivalent of the supply curve in economics. The corresponding demand (load) is modelled as inelastic, under the assumption that retail customers such as households and firms use electricity regardless of the wholesale price. The curves are then intersected to determine the market price. The two curves can change for each time step, depending on the available capacities, production, fuel prices and seasonality of load.

\begin{figure}[htbp]
	\centering
	\scalebox{1.35}{
	\includegraphics[width=0.75\textwidth]{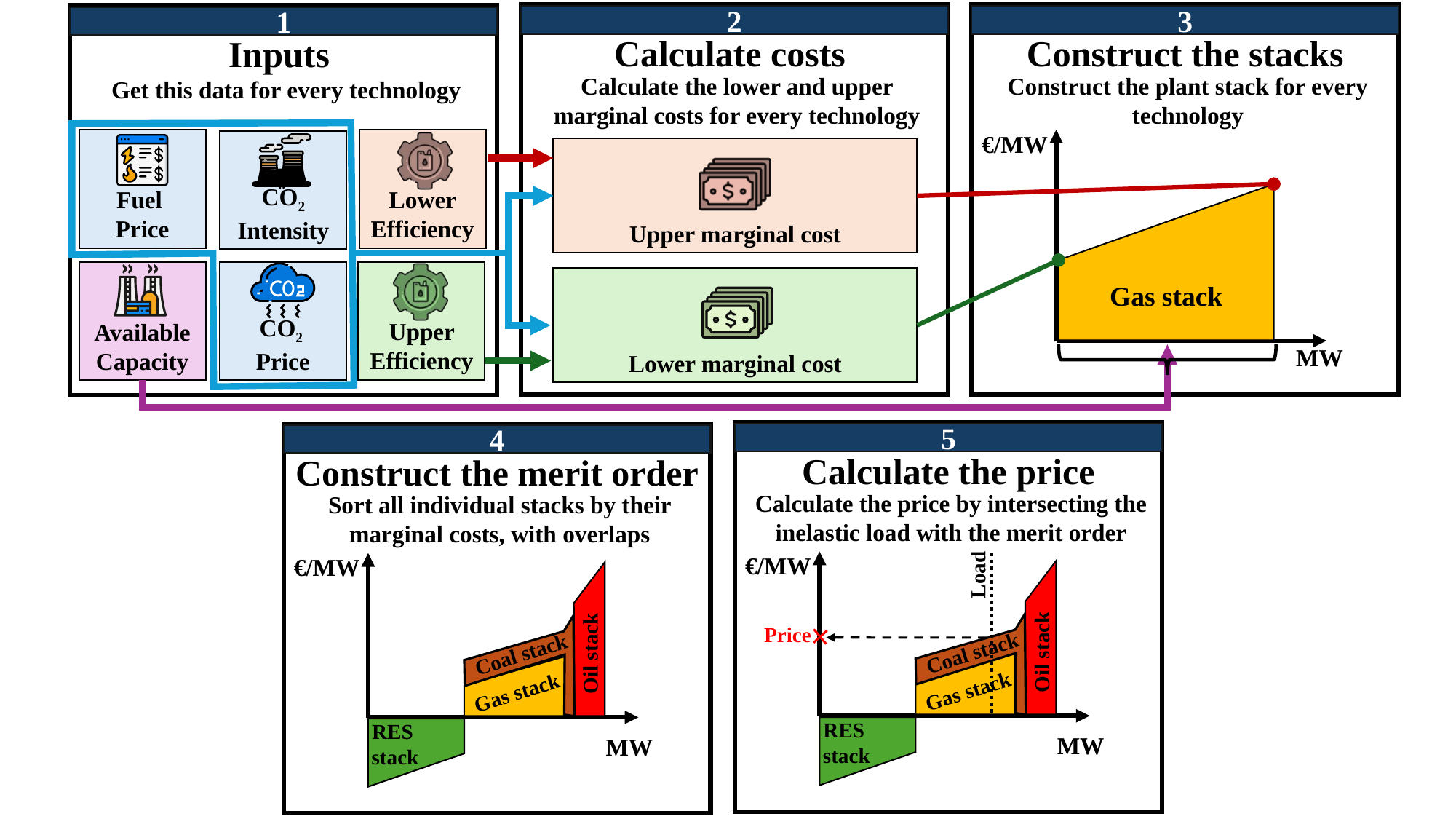}} 
	\caption{Constructing the merit order model.}
	\label{fig:mo_sketch}
\end{figure}

The model relies on three main components: the marginal costs of power plants, their available capacities and the load. The steps in constructing the merit order as summarized in Figure \ref{fig:mo_sketch} are the following:
\begin{enumerate}
	\item \textbf{Calculate marginal costs:} For each power plant type, calculate the marginal costs as: 
	\begin{align}
		\label{eq:var_costs}
		\text{VarCost}_{\text{pl},t}^{L}  & = \frac{\text{FuelPrice}_{\text{pl},t} + \varepsilon_{\text{pl}} \text{CO}_2\text{Price}_{t} }{\eta_{\text{pl}}^{L}} + \text{OtherCost}_{\text{pl},t},  \\
		\text{VarCost}_{\text{pl},t}^{U}  & = \frac{\text{FuelPrice}_{\text{pl},t} + \varepsilon_{\text{pl}} \text{CO}_2\text{Price}_{t} }{\eta_{\text{pl}}^{U}} + \text{OtherCost}_{\text{pl},t} \quad \forall \quad \text{pl} \in \mathcal{T} \nonumber
	\end{align}
	where $\text{pl}$ is the power plant type, $\eta_{\text{pl}}^{L}$ and $\eta_{\text{pl}}^{U}$ are the lower and upper efficiencies, $\varepsilon_{\text{pl}}$ is the $\text{CO}_2$ intensity factor. This assumes a linear cost curve where marginal costs are equal to variable costs. For the training phase the fuels will be the actual values while for the testing phase they will be the lagged values. The components, units and their estimates are summarized in Tables \ref{tab:var_costs} and \ref{tab:eff_co2}.

	For RES, $\text{FuelPrice}_{\text{pl},t}=\varepsilon_{\text{pl}}=0$ and $\eta_{\text{pl}}^{U}=\eta_{\text{pl}}^{U}=1$ simplifying Equation (\ref{eq:var_costs}) to:
	\begin{align}
		\text{VarCost}_{\text{pl},t}^{L} &= \underbrace{\text{OtherCost}_{\text{pl},t}}_{=b_{\text{pl},t}^{L}} \quad \forall \quad pl \in \mathcal{R}
		\label{eq:var_costs_res}
	\end{align}
	and equivalently for $\text{VarCost}_{\text{pl},t}^{U}$ where $b_{\text{pl},t}^{L}$ and $b_{\text{pl},t}^{U}$ (EUR/MWh) represents the lower and upper bidding prices, set as the negative absolute bounds of the subsidies received by the technology $\text{pl}$ (see Figure \ref{fig:intro_price_intervals}). Since RES have near-zero actual variable costs, the simplifying assumption $b_{\text{pl}} = 0$ is also commonly applied \citep{coester2018optimal}.

	\begin{table}[hbtp]
		\centering
		\scalebox{0.9}{
		\begin{tabular}{|l|l|l|}
			\hline
			$t$ & hours & time period \\
			$\text{pl}$ & - & power plant type \\
			$\textrm{MWh}_{el}$ & MWh & energy output (electric) \\
			$\textrm{MWh}_{th}$ & MWh & energy input (thermal) \\
			$\text{VarCost}_{\text{pl},t}$ & $ \text{EUR} / \text{MWh}_{el}$  & cost per MWh of electricity produced \\
			$\text{FuelPrice}_{\text{pl},t}$ & $ \text{EUR} / \text{MWh}_{th}$ & fuel price converted to $\text{EUR/MWh}_{}$ \\
			$\varepsilon_{\text{pl}}$ & $ tCO_2 / \text{MWh}_{th} $ & $\textrm{CO}_2$-intensity factor \\
			$\text{EUA}_{t}$ & $\text{EUR}/ tCO_2$ & $\textrm{CO}_2$ emissions allowance price \\
			$\eta_{\text{pl}}$ & $ \text{MWh}_{el}/ \text{MWh}_{th} $ & is the efficiency \\
			$\text{OtherCost}_{\text{pl},t}$ & $\text{EUR} / \text{MWh}_{el}$ & other variable costs \\
			\hline
		\end{tabular}
		}
		\caption{Components, units and description of Equation ({\ref{eq:var_costs}}).}
		\label{tab:var_costs}
	\end{table}


	\begin{table}[ht] 
		\centering
		\begin{tabular}{|l|ccccc|}
			\hline
			& Lignite & Coal & Gas & Oil & Nuclear \\
			\hline
			\textbf{Efficiency low}\footnotemark[1]  $\left ( \eta_{\text{pl}}^{\text{L}} \right )$ & 0.30 & 0.35 & 0.25 & 0.24 & 0.32 \\
			\textbf{Efficiency high}\footnotemark[1] $\left ( \eta_{\text{pl}}^{\text{U}} \right )$ & 0.43 & 0.46 & 0.40 & 0.44 & 0.42 \\
			\textbf{$\textrm{CO}_2$ intensity}\footnotemark[1] $\left ( \varepsilon_{\text{pl}} \right )$ & 0.40 & 0.30 & 0.20 & 0.30 & 0.03  \\
			\hline
		\end{tabular}
		\caption{Expert estimates for efficiencies and $\textrm{CO}_2$-intensity factors for plants.}
		\label{tab:eff_co2}
	\end{table}
	\footnotetext[1]{Expert estimates for efficiency and $\textrm{CO}2$-intensity factors as reported by \citep{beran2019modelling} and in the European Resource Adequacy Assessment (ERAA) report published by ENTSO-E. Link: https://www.entsoe.eu/outlooks/eraa/2023/eraa-downloads/.}

	\item \textbf{Get the available capacities:} For each power plant type, determine the available capacity at each time step, which is typically based on historical data or forecasts.
	\item \textbf{Construct individual stacks:} For each power plant type, construct the power plant stack, or individual merit order, as a trapezoid using linear interpolation between the lower and upper marginal costs, with the available capacity as the base. This is essentially a function that maps the produced volume or quantity $q$ (MWh) to a price $p$ (EUR/MWh):
	\begin{align}
	\label{eq:mo_indiv}
	& \text{MO}_{\text{pl},t}(q) = 
	\begin{cases}
		\begin{array}{ll}
		\text{VarCost}_{\text{pl},t}^{\text{L}} + \frac{q}{\text{Cap}_{\text{pl},t}} \left ( \text{VarCost}_{\text{pl},t}^{\text{U}} - \text{VarCost}_{\text{pl},t}^{\text{L}} \right ), &
		0 \le q \le \text{Cap}_{\text{pl},t}, \\
		\infty, & q > \text{Cap}_{\text{pl},t}
		\end{array}
	\end{cases}
\end{align}
where $\text{Cap}_{\text{pl},t}$ (MW) is the available capacity of the power plant type $\text{pl}$ at time $t$. For the training phase these will be the actual values while for the testing phase they will be the forecasted or lagged values.

	\item \textbf{Construct the supply curve:} Create a supply curve by plotting the cumulative available capacities against the sorted marginal costs, while taking into account overlaps. 
	
	\begin{figure}[hbtp]
	\centering
		\includegraphics[width=0.5\textwidth]{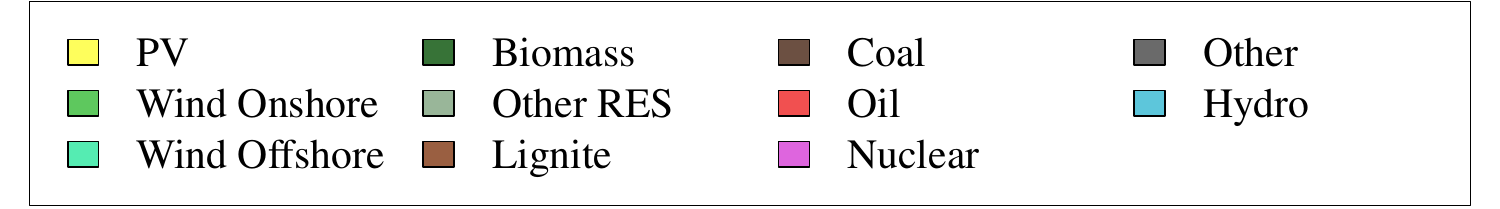}
	\linebreak
	\subfloat[Improper merit order]{
		\includegraphics[width=0.49\textwidth]{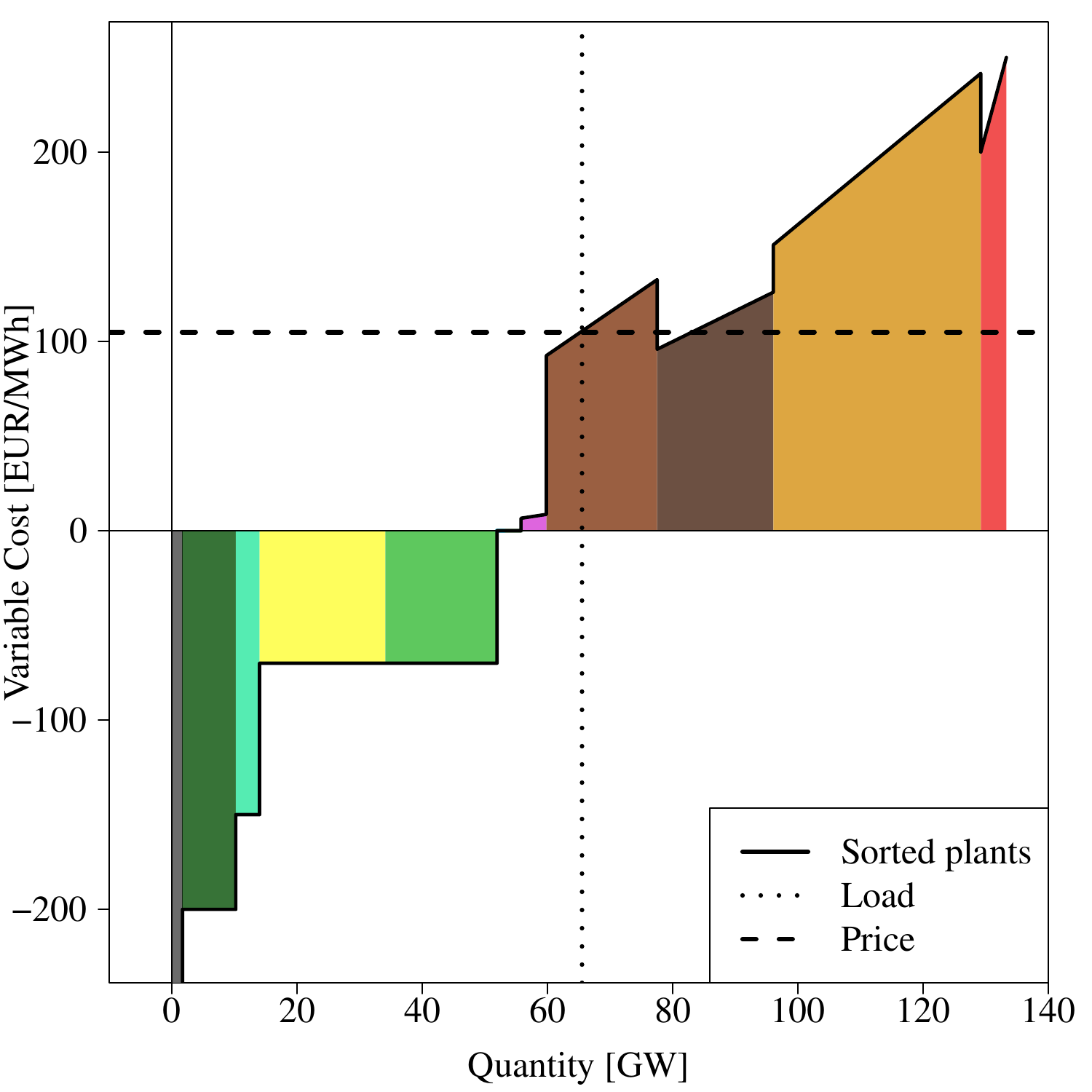}}
	\subfloat[Proper merit order]{
		\includegraphics[width=0.49\textwidth]{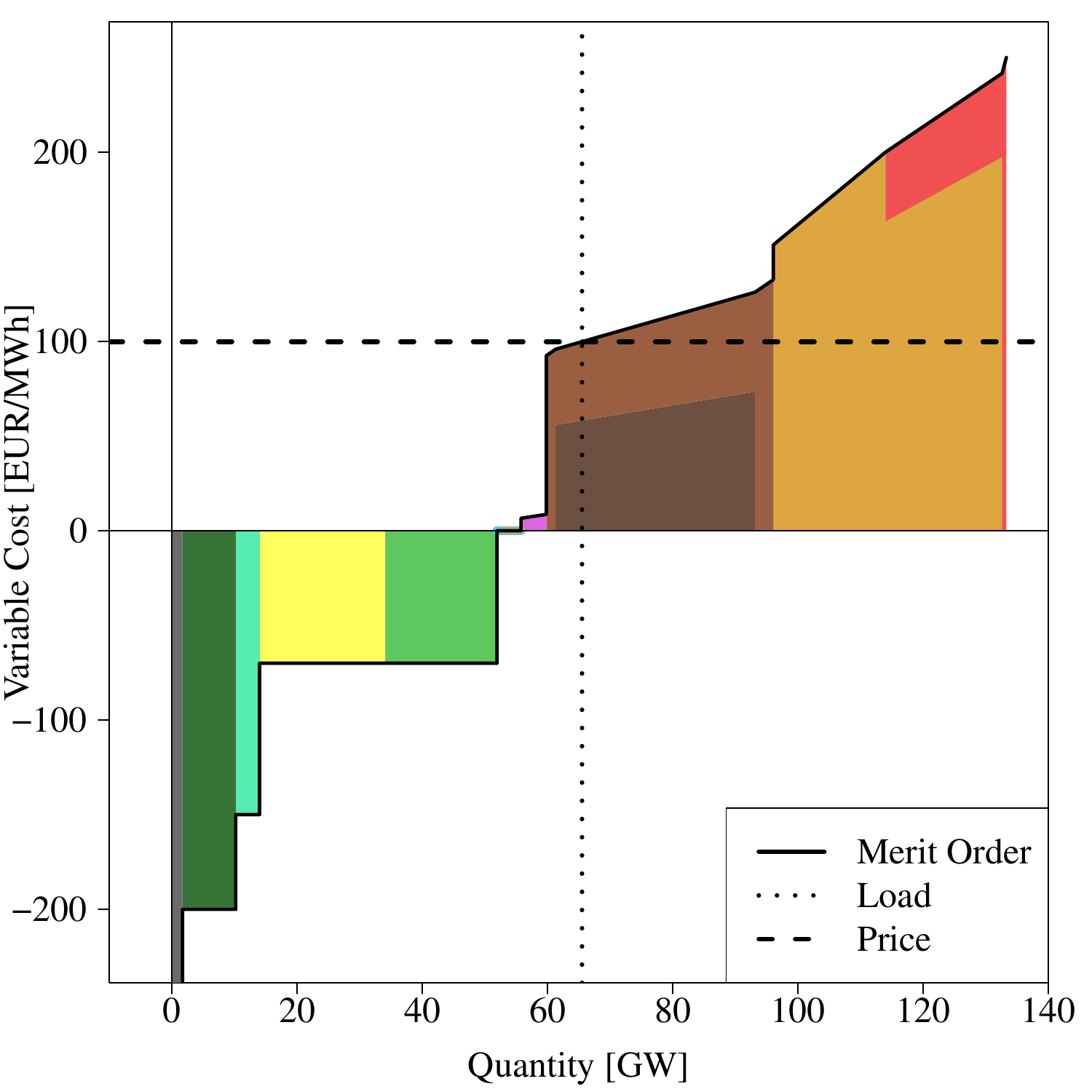}}
		\caption{Improper and proper merit order curves.}
	\label{fig:model_improper_proper_mo}
	\end{figure}
	
	Simply sorting by lower marginal costs does not account for overlaps, as shown on the right of Figure \ref{fig:model_improper_proper_mo}. To construct a proper merit order as on the right of Figure \ref{fig:model_improper_proper_mo}, the individual merit orders (stacks) must first be inverted. This allows them to be aggregated along the costs axis rather than the quantity axis:
	\begin{align}
		\label{eq:mo_inv}
		\text{MO}^{-1}_{\text{pl},t}(p) = \text{Cap}_{\text{pl},t} \frac{\max(\min(p, \text{VarCost}^{\text{U}}), \text{VarCost}^{\text{L}}) - \text{VarCost}^{\text{L}}_{\text{pl},t}}{\text{VarCost}^{\text{U}}_{\text{pl},t} - \text{VarCost}^{\text{L}}_{\text{pl},t}}
	\end{align}
	where $p$ is the price or cost in EUR/MWh.

	The system-wide inverse merit order is obtained by summing the inverse individual merit orders. Taking the inverse of this result produces the proper merit order:
	\begin{align}
		\label{eq:mo_proper}
		\text{MO}_{t}(q) = \left ( \sum_{\text{pl} \in \mathcal{T}} \text{MO}_{\text{pl},t}^{-1} \right )^{-1}(q)
	\end{align}

	\item \textbf{Intersect with demand:} Intersect the supply curve with the demand (load) to determine the market price and dispatched power plants.
	
	The resulting electricity price is given by the intersection of the merit order curve with the load at time $t$:
	\begin{align}
		\label{eq:mo_price}
		\text{Price}_{t} = \text{MO}_{t}(\text{Load}_t) + \epsilon_t,
	\end{align}
	as illustrated in Figure \ref{fig:mo_price_est}
	All power plant capacity to the left of the intersection point is dispatched, meaning that the power plants with marginal costs below the price are producing power. 
	To generate day-ahead price forecasts, all components in (\ref{eq:var_costs}) are replaced with forecasts for $t+1$ that are available at time $t$:
	\begin{align}
		\label{eq:mo_price_est}
		\widehat{\text{Price}}_{t+t} = \text{MO}_{t+1}(\widehat{\text{Load}}_{t+1}).
	\end{align}

\end{enumerate}

\begin{figure}[htbp]
    \centering
	\scalebox{0.9}{
	\includegraphics[width=1\textwidth]{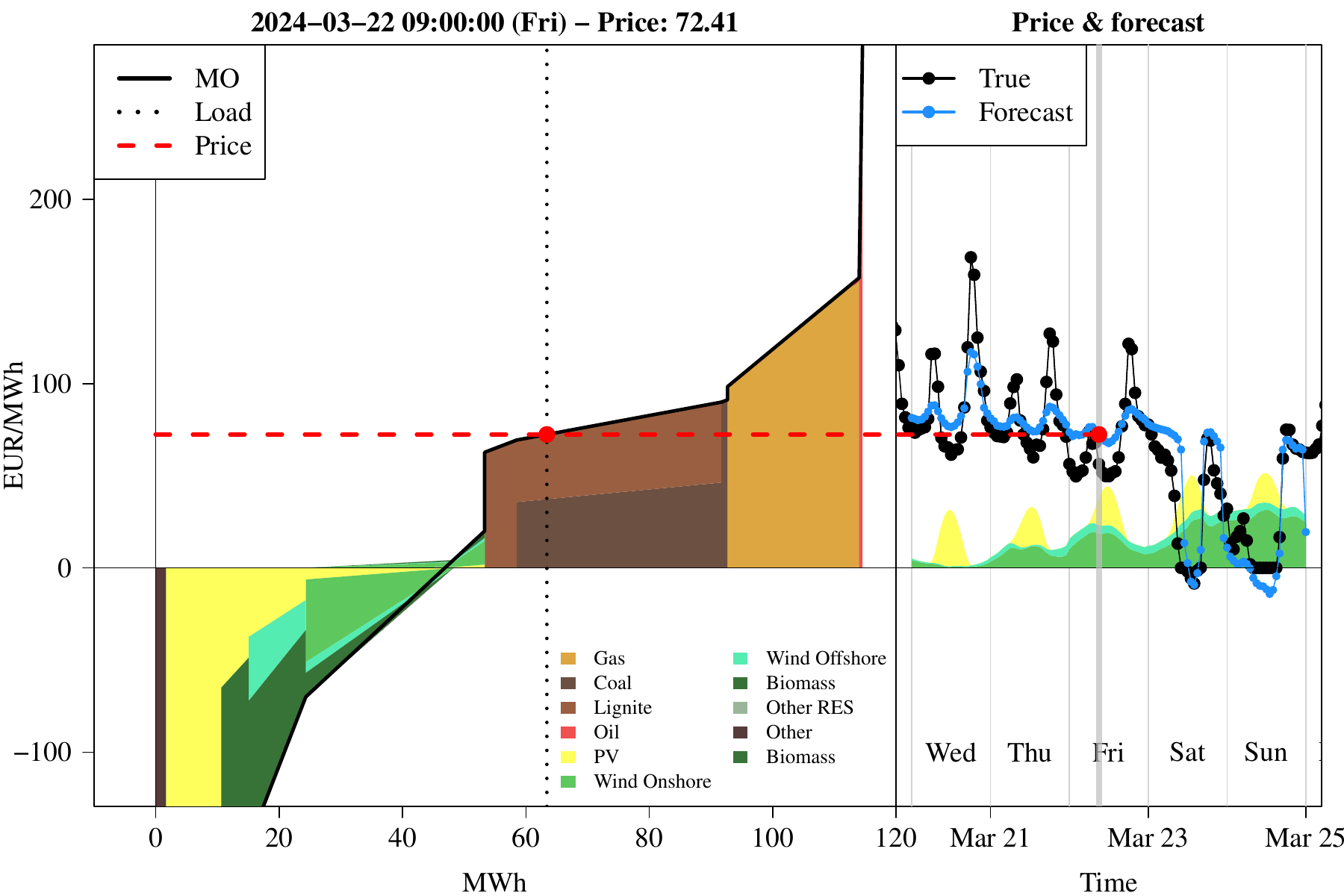}}
	\caption{Illustration of modelling the day-ahead price with the merit order.}
	\label{fig:mo_price_est}
\end{figure}

The individual merit order components, $\text{MOComp}_{\text{pl},t}(q)$, represented by the colored areas on the right side of Figure \ref{fig:model_improper_proper_mo}, are the marginal generation proportion for each plant type applied to the merit order curve. For each plant type, this is calculated by multiplying the price for each segment with the ratio of its generated power to total generated power.
In other words, for each plant type the ratio of the power quantity produced is multiplied with the price for each segment where the overlaps change. 
\begin{align}
	\label{eq:mo_comp}
	\text{MOComp}_{\text{pl},t}(q) = \frac{\text{d} \text{MO}^{-1}_{\text{pl},t}(\text{MO}_{t}(q))}{\text{d}\text{MO}^{-1}_{t}(\text{MO}_{t}(q))}\text{MO}_{t}(q) 
\end{align}
where the fraction is the ratio of the differential of the whole inverse merit order to the differential of the individual merit order for a certain price. The fraction is not the derivative. The differentials $\text{d} \text{MO}_t^{-1}(p)$ and $\text{d} \text{MO}_{\text{pl},t}^{-1}(p)$ are infinitesimal differences in quantity when the price changes by a very small amount $\text{d} p$. 

In the linear case, it is sufficient to compute the differences between the bounds of each linear segment of the merit order, with interpolation between them (see right hand of Figure \ref{fig:model_improper_proper_mo}):
\begin{align}
	\label{eq:mo_comp_lin}
	\text{MOComp}_{\text{pl},t}(q_i) = \frac{\text{MO}^{-1}_{\text{pl},t}(p_{i})-\text{MO}^{-1}_{\text{pl},t}(p_{i-1})}{\text{MO}^{-1}_{t}(p_{i})-\text{MO}^{-1}_{t}(p_{i-1})}\text{MO}_{t}(q_i),
\end{align} 
where $p_i \in (\mathcal{P}, \le)$ represents the ordered set of all lower and upper variable costs for all technologies, defined as $\mathcal{P} = \{ (\text{VarCost}_{\text{pl},t}^{\text{L}})_{\text{pl} \in \mathcal{T}}, (\text{VarCost}_{\text{pl},t}^{\text{U}})_{\text{pl} \in \mathcal{T}}\}$, and $q_i = \text{MO}_t^{-1}(p_i)$ is the quantity corresponding to $p_i$.

\subsection{Data-driven merit order}
\label{sec:mo_model_optim}
We introduce the \emph{data-driven merit order} as a generalization of (\ref{eq:mo_price}). For this we first define the volume space $\mathbb{Q} = [0, \infty )$ as the set of all possible quantities or volumes $q$ produced in MWh, and the price space $\mathbb{P}= [-500,3000]$\footnote{The market operator can adjust the upper and lower price bounds, as stated in Art. 41(1) of Commission Regulation (EU) 2015/1222 of 24th July 2015 (CACM Regulation). In Germany, the maximum price day-ahead price was raised from 3000 EUR/MWh to 4000 EUR/MWh on 10 May 2022, as announced by EpexSpot on 11 April 2022. Link: https://www.epexspot.com/en/news/harmonised-maximum-clearing-price-sdac-be-set-4000-eurmwh-10th-may-2022} as the set of all possible prices $p$ in EUR/MWh. Then, the data-driven merit order is a function which maps the domain (input-space) $\mathbb{Q}$ to the range (output-space) $\mathbb{P}$:
\begin{align}
	\text{MO}_t(q;\Theta): \mathbb{Q} \mapsto  \mathbb{P}
\end{align}
where $\Theta$ is the set of optimizable input parameters:
\begin{align}
	\Theta = \left \{
		(\eta_{\text{pl}}^{\text{L}})_{\text{pl} \in \mathcal{C}}, 
		(\eta_{\text{pl}}^{\text{U}})_{\text{pl} \in \mathcal{C}},
		(b^{\text{{L}}}_{\text{pl}})_{\text{pl}\in \mathcal{R}}, 
		(b^{\text{U}}_{\text{pl}})_{\text{pl}\in \mathcal{R}}\right \}, \nonumber
\end{align}
In contrast to the classical merit order (\ref{eq:mo_proper}), these parameters are not fixed but rather given as an additional.
All other inputs, such as the capacities, fuel prices, and $\text{CO}_2$ intensities, are assumed to be exogenous to the model and are held fixed. 
Thus, the classical merit order model (\ref{eq:mo_proper}) is a special case of $\text{MO}_t(q;\Theta)$:
\begin{align}
	\text{MO}_t(q;\Theta_{\text{init}}) = \text{MO}_{t}(q).
\end{align}
where $\Theta_{\text{init}}$ represents the expert estimates from Table \ref{tab:eff_co2}.

\subsection{Estimation}
The optimal parameter set $\hat{\Theta}$ is determined by solving:
\begin{align}
	\label{eq:mo_optim}
	& \hat{\Theta} = \underset{\Theta}{\arg \min} \left ( \frac{1}{N} \sum_{t=1}^{N} \left | \text{MO}_t(\text{Load}_t; \Theta) - \text{Price}_t \right | \right ) \\
	& \text{s.t.} \quad \Theta^{\text{L}}  \le \Theta \le \Theta^{\text{U}} \nonumber
\end{align}
which minimizes the mean absolute error (MAE) between the modelled price $\text{MO}_t(q_t; \Theta) $ and the observed price $\text{Price}_t$, where $q_t = \text{Load}_t$. The parameter search space in constrained by the lower and upper bounds $\Theta^{\text{L}}$ and $ \Theta^{\text{U}}$, as summarized in Table \ref{tab:theta_bounds}. 
The initial starting point for the optimization $\Theta_{\text{init}}$ lies within the search space and corresponds to the parameters of the classical merit order model. Consequently, the classical merit order model is an embedded within this framework. This means that, if the expert assumptions of the classical merit order model are indeed optimal, then the optimization will return these same parameters.   

Fundamental power plant parameters in $\Theta$ are assumed to be constant over time, hence they do not depend on the time index $t$. They are solved once for the training period and held constant for the test period. The optimization problem is non-convex and involves multivariate nonlinear parameter estimation, making it a black-box optimization problem. To solve this problem, we use the R package \textit{mlrMBO} \citep{mlrMBO}, which does not require a gradient. Each optimization run was configured to execute for 60 minutes.

\subsection{Extensions}
The function itself $\text{MO}_t(q;\Theta)$ and the parameter set $\Theta$ can be extended to include other modeling aspects. We introduce these extensions below and collect them in $\Theta_{\text{ext}}$:
\begin{align}
	\label{eq:theta_ext}
	&\Theta_{\text{ext}} = \left \{
		(\eta_{\text{pl}}^{\text{L}})_{\text{pl} \in \mathcal{C}}, 
		(\eta_{\text{pl}}^{\text{U}})_{\text{pl} \in \mathcal{C}}, 
		(b^{\text{{L}}}_{\text{pl}})_{\text{pl}\in \mathcal{R}}, 
		(b^{\text{U}}_{\text{pl}})_{\text{pl}\in \mathcal{R}},
		(\text{cf}_{\text{pl}})_{\text{pl} \in \mathcal{T}_{\text{ext}}},
		(\text{mr}_{\text{pl}})_{\text{pl} \in \mathcal{T}_{\text{ext}}},
		\text{gs} \right \}, \\
	&\mathcal{T}_{\text{ext}} = \mathcal{T} \cup \{ \text{hydro}, \text{net import}\}  \nonumber 
\end{align}
where the additional components are detailed below. These parameters are estimated using the same procedure as in (\ref{eq:mo_optim}).

\subsubsection{Capacity correction factors $(\textrm{cf}_{\textrm{pl}})$}
Reported plant capacities (Figure \ref{fig:data_cap_gen}) are systematically underreported, as only plants above 100MW must disclose data on capacity, outages, and generation\footnote{As required by Commission Regulation (EU) No 543/2013 of 14 June 2013.}.
To account for this, we apply a correction factor $1 \le \text{cf}_{\text{pl}} \le 2$ by multiplying it with the reported available capacity $\text{Cap}_{\text{pl},t}$ of non-intermittent power plants where capacities less than 100MW are common. This assumes similar maintenance behavior across reported and unreported plants\footnote{Alternatively, $\text{cf}_{\text{pl}}$ could be added to $\text{Cap}_{\text{pl},t}$, thus the type of operator could be treated as a parameter to be optimized.}.

\subsubsection{Must-run stack $(\textbf{mr}_{\text{pl}})$}
Figures \ref{fig:intro_price_intervals} and \ref{fig:data_cap_gen} show that many plant types sustain a minimum generation level, never dropping to zero. This \textit{must-run} behavior reflects plants operating continuously, regardless of market signals. Contributing factors include long-term contracts, high startup and ramping costs which can exceed losses incurred from negative price periods, or combined heat and power (CHP) obligations.
We capture this with a must-run share  $0 \le \text{mr}_{\text{pl}} \le 1$ for each plausible plant type. The must-run capacities are subtracted from the individual available capacities and aggregated system-wide: 
\begin{align}
\text{MustRun}_t = \sum_{\text{pl}} \text{mr}_{\text{pl}} \cdot \text{Cap}_{\text{pl,t}}
\end{align}
It is represented as a virtual plant with a marginal cost of $-500$ EUR/MWh, ensuring dispatch guarantee.

\subsubsection{Gas stack split $(\text{gs})$}
Gas power plants are highly flexible and traditionally among the most expensive conventional sources after oil. They are typically divided into combined cycle gas turbines (CCGT), which are more efficient, and open cycle gas turbines (OCGT), used mainly for peak loads.
To reflect this in the merit order model, we introduce a gas stack split factor  $0 \le \text{gs} \le 1$, defining the share of CCGT and OCGT capacity:
\begin{align}
& \text{Cap}^{\text{new}}_{\text{gas},t} = \text{gs} \cdot \text{Cap}_{\text{gas},t}, \
& \text{Cap}^{\text{new}}_{\text{gas2},t} = (1 - \text{gs}) \cdot \text{Cap}_{\text{gas},t}. \nonumber
\end{align}
This introduces two gas plant types, each with distinct parameters, increasing model flexibility. The optimal split is optimized. The same approach can be extended to other technologies, for example splitting RES by subsidy regime.

\subsubsection{Hydro power forecast}
Hydropower in Germany includes run-of-river, reservoir, and pumped storage plants. As pumped storage involves both consumption and generation for temporal price arbitrage purposes, it is excluded from the merit order model to avoid such complexities.
Thus, our model considers only run-of-river and reservoir output (Figure \ref{fig:data_cap_gen}). Generation is forecasted using a LASSO-estimated model incorporating autoregressive terms, cross-hour dependencies, weekday effects, and seasonal patterns (see Appendix \ref{eq:hydro}). Hydropower is then added to the merit order model as an additional technology type similar to RES.

\subsubsection{Net import forecast}
Although classical merit order models exclude cross-border flows, net imports significantly affect Germany's power balance (Figure \ref{fig:data_cap_gen}). We define net imports as:
\begin{align}
	\text{NetImport}_t^{\text{DE}} = \sum_{\text{bzn} \in \mathcal{B}_{\text{DE}}}\text{Import}_t^{\text{bzn}} - \sum_{\text{bzn} \in \mathcal{B}_{\text{DE}}}\text{Export}_t^{\text{bzn}}
\end{align}
where $t$ denotes the hour, and $\mathcal{B}_{\text{DE}}$ is the set of all bidding zones with physical power flows to and from Germany. 
Net imports are forecasted using a LASSO-estimated model with autoregressive terms, cross-hour effects, and weekly and annual seasonality (see Appendix \ref{eq:net_import}). In the merit order model, net imports are treated as a virtual plant with a marginal cost of $-500$ EUR/MWh, implicitly assuming that Germany imports when foreign prices are lower. This shifts the supply curve to the right for net imports or to the left for net exports.

\section{Results}
\label{sec:result}
\subsection{Forecasting study design and computation time}
To evaluate the proposed model, we follow a standard data-driven forecasting setup. The data is split into a training set of 1 year (01/10/2022–01/10/2023) for parameter estimation as in (\ref{eq:mo_optim}), and a test set of 1 year (01/10/2023–01/10/2024) for out-of-sample evaluation. The parameters $\Theta$ are estimated using the training set and then fixed for the test set.
During training, actual RES generation is used. In the test phase, RES values are replaced with day-ahead forecasts. 

The merit order model was implemented in \textit{C++} and interfaced with \textit{R} via \textit{Rcpp} R package \citep{rcpp}. It is highly efficient, requiring approximately 1 second per run to simulate a full year of hourly prices. Table \ref{tab:mo_computation_time} reports computation time statistics using the \textit{microbenchmark} R package \citep{microbenchmark} on a Linux machine (64 GB RAM, 14-core 12th Gen Intel® Core™ i7-12700H).
\begin{table}[ht]
	\centering
	\scalebox{0.8}{
	\begin{tabular}{|l|ccccccc|}
		\hline
		& \bf Min & \bf LQ & \bf Mean & \bf Median & \bf UQ & \bf Max & \bf Neval \\ 
		\hline
		\bf  Model run time (s) & 0.41 & 0.58 & 0.74 & 0.71 & 0.88 & 1.45 & 100 \\ 
		\hline	
	\end{tabular}}
	\caption{Computation time statistics of a single run of the extended merit order model. One run calculates prices for 8760 hours and for 14 technology types.} 
	\label{tab:mo_computation_time}
\end{table}
Training used Bayesian stochastic optimization via the \textit{mlrMBO} R package with a 60-minute time limit. Around 3600 parameter configurations were evaluated per run, with the best (lowest training error) retained. 

\subsection{Benchmarks and results}

To assess the \textbf{data-driven merit order} model, we compare it against leading data-driven forecasting methods: \textbf{Lasso}, \textbf{XGBoost}, and a \textbf{Neural Network}, all trained on the same data as the merit order model, including autoregressive and cross-hours effects (see Appendix \ref{eq:expert}), and evaluated using a rolling window of 1 year over the test period. A \textbf{Naive} persistence model (see Appendix \ref{eq:naive}) serves as a baseline. Results are reported in Table \ref{tab:errors}. The optimal parameter values, bounds and initial values for the merit order model are summarized in Appendix Table \ref{tab:theta_bounds}.

\begin{table}[ht]
\centering
	\scalebox{0.8}{
	\begin{tabular}{|C{2cm}|C{2cm}|C{2cm}|C{2cm}|C{2cm}|C{2.3cm}|C{2cm}|}
		\hline
		\bf Model & \bf Naive & \bf Lasso & \bf XGBoost & \bf NNet & \bf Classic MO & \bf Data MO \\ 
		\hline
		Error & \cellcolor[rgb]{1,0.6,0.6} $\text{24.96}$ & \cellcolor[rgb]{0.678,1,0.6} $\text{13.11}$ & \cellcolor[rgb]{0.73,1,0.6} $\text{13.96}$ & \cellcolor[rgb]{1,0.99,0.6} $\text{18.56}$ & \cellcolor[rgb]{0.876,1,0.6} $\text{16.36}$ & \cellcolor[rgb]{0.6,1,0.6} $\textbf{\text{11.83}}$ \\ 
		\hline
		Skill & \cellcolor[rgb]{1,0.6,0.6} $\text{100}$\% & \cellcolor[rgb]{0.691,1,0.6} $\text{53}$\% & \cellcolor[rgb]{0.736,1,0.6} $\text{56}$\% & \cellcolor[rgb]{1,0.992,0.6} $\text{74}$\% & \cellcolor[rgb]{0.887,1,0.6} $\text{66}$\% & \cellcolor[rgb]{0.6,1,0.6} $\textbf{\text{47}}$\% \\ 
		\hline
	\end{tabular}}
	\caption{Test MAE (EUR/MWh) and skill improvement compared to the \textit{Naive} model for considered models.} 
	\label{tab:errors}
\end{table}

The proposed \textbf{data-driven merit order} achieves the lowest MAE, improving by $1.2$EUR/MWh ($\approx 10\%$) over the best machine learning benchmark, \textbf{Lasso}. Notably, this result is obtained without using autoregressive price terms, relying solely on an optimized fundamental framework, unlike the machine-learning benchmarks, which include autoregressive components (see Appendix (\ref{eq:expert})).

Figure \ref{fig:result_mo_optim}.a shows that the optimized merit order has a more complex shape than the classical version (Figure \ref{fig:mo_price_est}), featuring exponential tails and a flat middle section. This structure mitigates the issues of the classical model, improves the representation of price volatility at both extremes, and captures clustering around zero due to the flat segment (Figure \ref{fig:result_mo_optim}.b).

\begin{figure}[hbtp]
	\centering
		\includegraphics[width=0.6\textwidth]{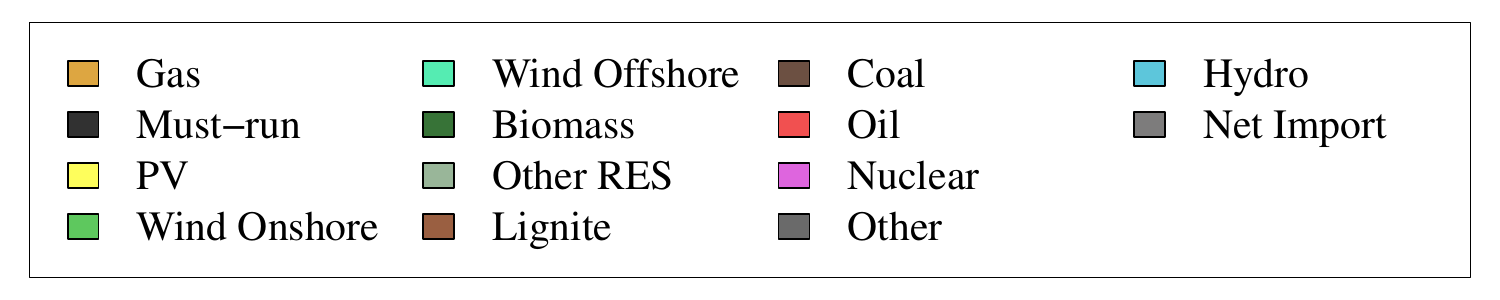}
	\linebreak
    \subfloat[Data-driven MO model]{
        \includegraphics[width=0.49\textwidth]{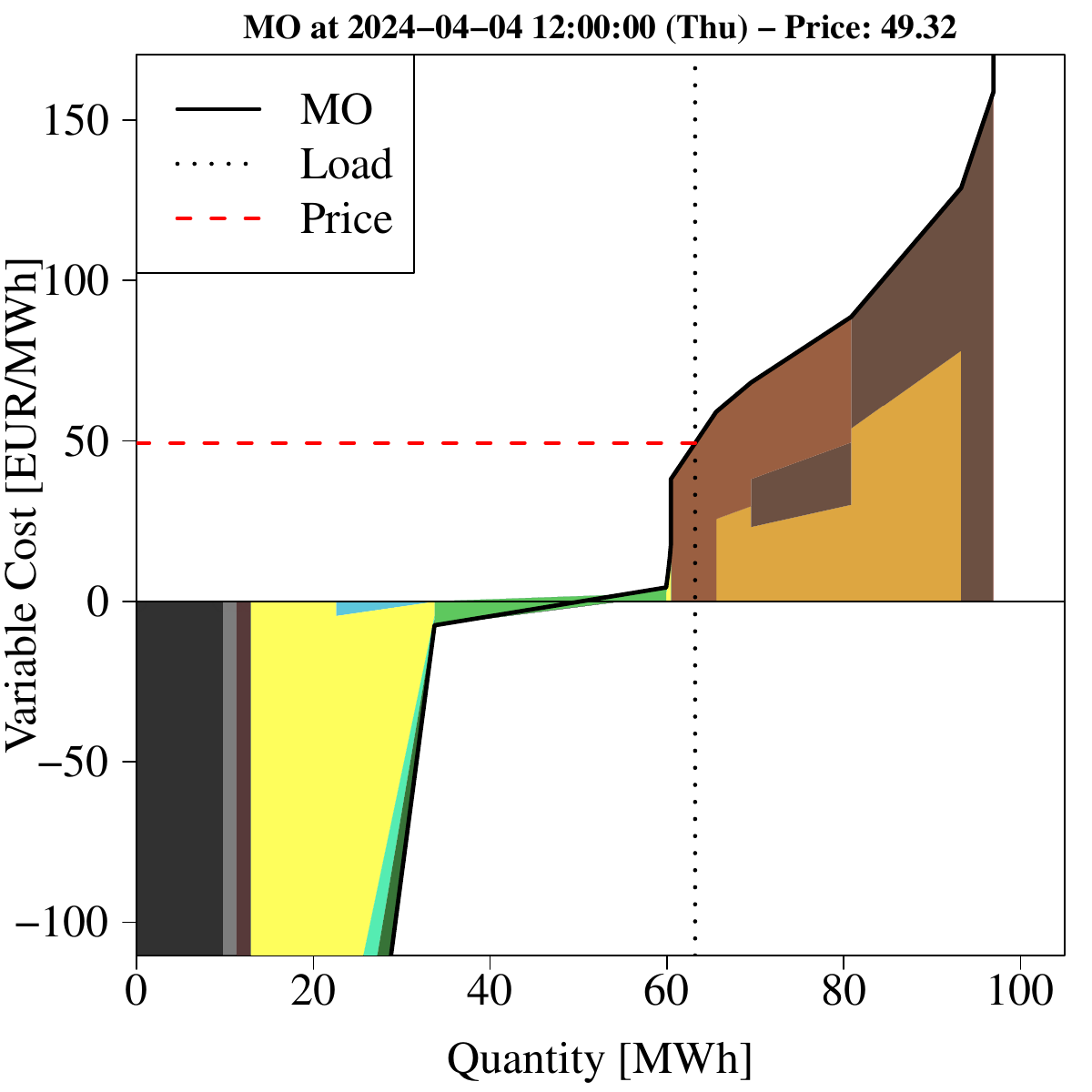}}
    \subfloat[Forecasts comparisons (2024)]{
        \includegraphics[width=0.49\textwidth]{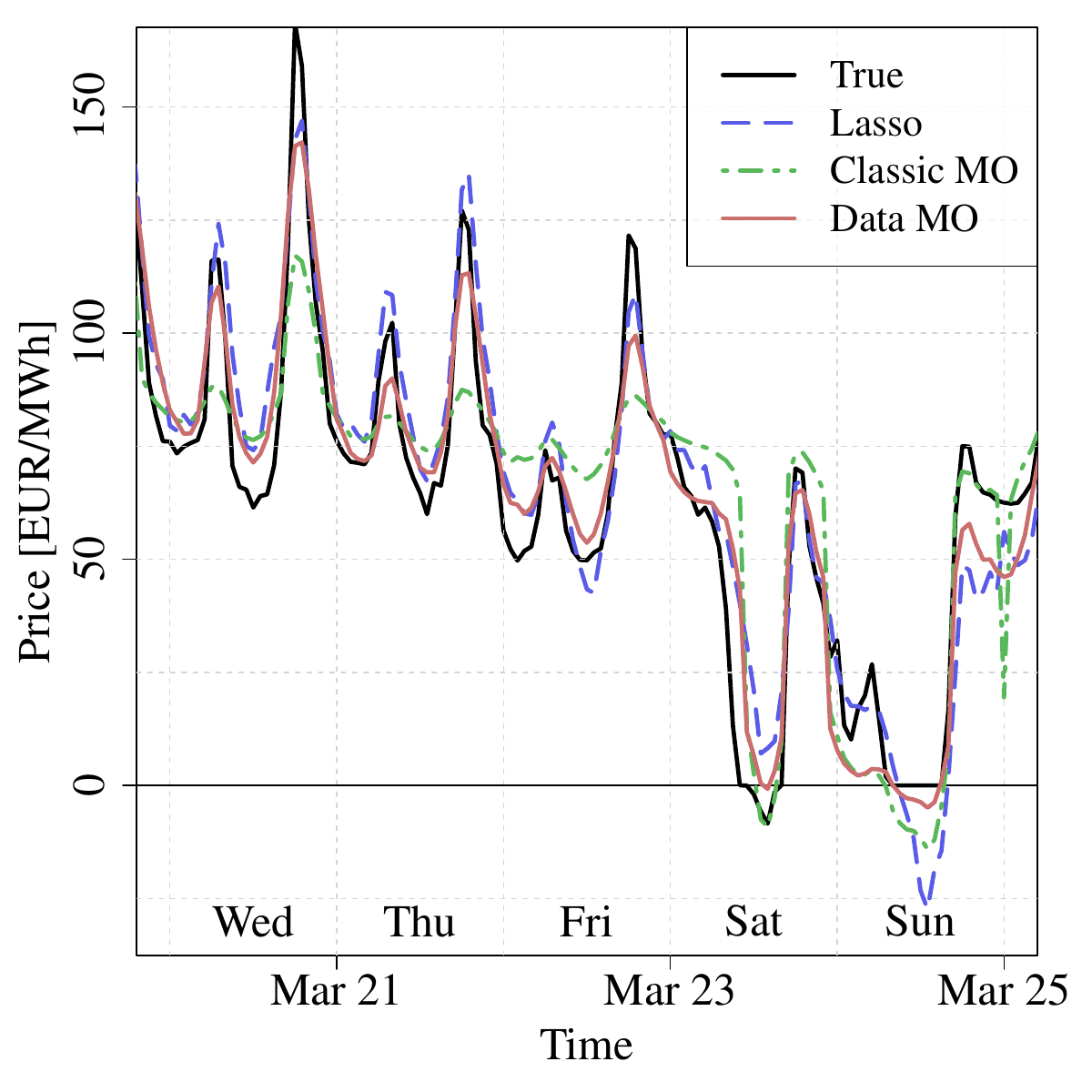}}
	\caption{Visualization of a resulting data-driven merit order and forecasts.}
	\label{fig:result_mo_optim}
\end{figure}

\subsection{Model breakdown}
To evaluate the contribution of each model extension to forecast accuracy, we perform optimizations for all $2^6 = 64$ combinations of the six parameter types. Using a greedy forward selection procedure, we begin with single-parameter optimizations and iteratively add parameters, selecting the combination with the lowest training error at each step. This process is illustrated in Figure \ref{fig:forward_tree}, where the forward selection path is highlighted in grey and the optimal test-based \textbf{data-driven merit order} configuration is highlighted in green.

\begin{figure}[htpb]
	\centering
\begin{center}
        \scalebox{0.58}{
    \begin{tikzpicture}[
        neutralnode/.style={rectangle, draw=black, very thick, minimum width = 30mm,  minimum height = 23mm, align = center,},
        chosennode/.style={rectangle, draw=black, very thick, minimum width = 1mm,  minimum height = 1mm, align = center, anchor = base north}
        ]

    \definecolor{chosen}{rgb}{0.7216, 0.7137, 0.7137}
    \definecolor{best}{rgb}{0.6549, 0.9451, 0.6549} 
	\definecolor{title}{rgb}{0.8235, 0.8235, 0.8235} 
	\definecolor{teal}{rgb}{0.4, 0.8, 0.8}

 \node[neutralnode, xshift=0cm,fill = chosen,line width=2pt] (M11)  at (5,0) {\setlength\extrarowheight{-6pt} \begin{tabular}{c} \bf Efficiencies\\Train: 23.38\\Test: 13.88\end{tabular}};
\node[neutralnode, right=of M11] (M12) {\setlength\extrarowheight{-6pt} \begin{tabular}{c} \bf Cap Factors\\Train: 30.35\\Test: 17.73\end{tabular}};
\node[neutralnode, right=of M12] (M13) {\setlength\extrarowheight{-6pt} \begin{tabular}{c} \bf Tech Split\\Train: 35.35\\Test: 17.04\end{tabular}};
\node[neutralnode, right=of M13] (M14) {\setlength\extrarowheight{-6pt} \begin{tabular}{c} \bf Must Run\\Train: 35.47\\Test: 16.36\end{tabular}};
\node[neutralnode, right=of M14] (M15) {\setlength\extrarowheight{-6pt} \begin{tabular}{c} \bf Net Import\\Train: 39.54\\Test: 17.94\end{tabular}};
\node[neutralnode, right=of M15] (M16) {\setlength\extrarowheight{-6pt} \begin{tabular}{c} \bf Hydro\\Train: 36.1\\Test: 16.49\end{tabular}};
\node[neutralnode, xshift=2cm,fill = chosen,line width=2pt] (M21) at (5,-3) {\setlength\extrarowheight{-6pt} \begin{tabular}{c}+ \\ \bf Cap Factors \\Train: 23.11 \\Test: 16.12\end{tabular}};
\node[neutralnode, right=of M21] (M22) {\setlength\extrarowheight{-6pt} \begin{tabular}{c}+ \\ \bf Tech Split \\Train: 23.28 \\Test: 15.28\end{tabular}};
\node[neutralnode, right=of M22] (M23) {\setlength\extrarowheight{-6pt} \begin{tabular}{c}+ \\ \bf Must Run \\Train: 28.21 \\Test: 17.99\end{tabular}};
\node[neutralnode, right=of M23] (M24) {\setlength\extrarowheight{-6pt} \begin{tabular}{c}+ \\ \bf Net Import \\Train: 24.1 \\Test: 13.67\end{tabular}};
\node[neutralnode, right=of M24] (M25) {\setlength\extrarowheight{-6pt} \begin{tabular}{c}+ \\ \bf Hydro \\Train: 23.29 \\Test: 13.38\end{tabular}};
\node[neutralnode, xshift=4cm] (M31) at (5,-6) {\setlength\extrarowheight{-6pt} \begin{tabular}{c}+ \\ \bf Tech Split \\Train: 23.17 \\Test: 13.89\end{tabular}};
\node[neutralnode, right=of M31] (M32) {\setlength\extrarowheight{-6pt} \begin{tabular}{c}+ \\ \bf Must Run \\Train: 28.52 \\Test: 25.74\end{tabular}};
\node[neutralnode,fill = chosen,line width=2pt, right=of M32] (M33) {\setlength\extrarowheight{-6pt} \begin{tabular}{c}+ \\ \bf Net Import \\Train: 22.43 \\Test: 12.62\end{tabular}};
\node[neutralnode, right=of M33] (M34) {\setlength\extrarowheight{-6pt} \begin{tabular}{c}+ \\ \bf Hydro \\Train: 24.08 \\Test: 15.08\end{tabular}};
\node[neutralnode, xshift=6cm,fill = chosen,line width=2pt] (M41) at (5,-9) {\setlength\extrarowheight{-6pt} \begin{tabular}{c}+ \\ \bf Tech Split \\Train: 22.04 \\Test: 12.87\end{tabular}};
\node[neutralnode, right=of M41] (M42) {\setlength\extrarowheight{-6pt} \begin{tabular}{c}+ \\ \bf Must Run \\Train: 24.18 \\Test: 15.18\end{tabular}};
\node[neutralnode, right=of M42] (M43) {\setlength\extrarowheight{-6pt} \begin{tabular}{c}+ \\ \bf Hydro \\Train: 22.2 \\Test: 12.51\end{tabular}};
\node[neutralnode, xshift=8cm] (M51) at (5,-12) {\setlength\extrarowheight{-6pt} \begin{tabular}{c}+ \\ \bf Must Run \\Train: 25.73 \\Test: 15.95\end{tabular}};
\node[neutralnode,fill = best,line width=2pt, right=of M51] (M52) {\setlength\extrarowheight{-6pt} \begin{tabular}{c}+ \\ \bf Hydro \\Train: 21.63 \\Test: 11.83\end{tabular}};
\node[neutralnode,fill = chosen, xshift=10cm,line width=2pt] (M61) at (5,-15) {\setlength\extrarowheight{-6pt} \begin{tabular}{c} + \\ \bf Must Run \\Train: 23.17 \\Test: 14.86\end{tabular}};
%
\path let
  \p1 = ($(M11)!.5!(M16)$) 
in
  coordinate (CenterTop) at (\x1, 3); 
\node[neutralnode, yshift = 1cm, minimum height = 40mm] (M0) at (CenterTop) 
  {\setlength\extrarowheight{-6pt} \begin{tabular}{c} \bf Classic MO \\ Train: 35.47 \\ Test: 16.36\end{tabular}};

\node[neutralnode, fill = teal, left=of M11, yshift = -215, minimum width = 10mm, minimum height = 175mm] (MOtext) {\rotatebox{90}{ \large \bf Optimization}};
\node[neutralnode, fill = title, above=of MOtext, minimum width = 10mm, minimum height = 40mm] (MOclassic) {\rotatebox{90}{\large \bf No Optimization}};

 \foreach \x in {M11,M12,M13,M14,M15,M16} {
            \draw[->, thick] (M0.south)  -- (\x.north);
            } 
 \foreach \x in {M21,M22,M23,M24,M25} {
            \draw[->, thick] (M11.south)  -- (\x.north);
            } 
 \foreach \x in {M31,M32,M33,M34} {
            \draw[->, thick] (M21.south)  -- (\x.north);
            } 
 \foreach \x in {M41,M42,M43} {
            \draw[->, thick] (M33.south)  -- (\x.north);
            } 
 \foreach \x in {M51,M52} {
            \draw[->, thick] (M41.south)  -- (\x.north);
            } 
 \foreach \x in {M61} {
        \draw[->, thick] (M52.south)  -- (\x.north);
        } 
        \end{tikzpicture}}
        \end{center}
\caption{Forward selection tree of nested MO models based on train MAE. Each row adds one variable to be optimized. Grey boxes indicate the optimal path. The final row is the full model with all extensions. The green box marks the lowest test MAE achieved.}
\label{fig:forward_tree}
\end{figure}

After efficiencies, the most impactful extensions are joint optimization of capacity factors, net imports, gas split, and hydro. The must-run extension appears to add little additional value. 
While training error generally decreases with each added parameter, though not guaranteed due to the stochastic optimization, the test error does not always follow the same trend. This highlights the model's sensitivity to the calibration window and potential overfitting.

To further evaluate model performance, Table \ref{tab:errors_by_res} reports the MAE across quantiles of residual load, defined as day-ahead load minus day-ahead renewable generation: $\text{ResLoad}_t = \text{Load}_t - (\text{PV}_t + \text{WindOnshore}_t + \text{WindOffshore}_t)$. This represents the portion of demand that must be met by conventional power plants.
 Since high levels of residual load often coincides with price spikes and low levels with low or negative prices (see Figure \ref{fig:intro_price_intervals}), upper quantiles reflect high-price periods, while lower quantiles reflect low-price conditions. This residual load-based binning is preferred over price-based binning due to structural price shifts following the 2022 European energy crisis.

The \textbf{data-driven merit order} and its variants outperform nearly all benchmarks across all quantiles, except in the upper range, where \textbf{Lasso} performs better on two occasions. This is likely to be due to its autoregressive component capturing price peaks. Nonetheless, the merit order model is competitive in these cases and even surpasses \textbf{Lasso} at the $65\%$ quantile. It performs best across most quantiles, clearly outperforming all other models.

\begin{table}[ht]
\centering
\resizebox{0.95\linewidth}{!}{
\scalebox{0.6}{
\setlength\extrarowheight{-4pt}	
\begin{tabular}{|
	>{\centering\arraybackslash}p{2cm}|
	>{\centering\arraybackslash}p{1cm}|
	*{10}{>{\centering\arraybackslash}p{1cm}|}}
\hline
\multirow{2}{*}{
	{\rotatebox{0}{\parbox[b][4.5cm][b]{3cm}{\begin{tabular}{c}
    \textbf{Residual} \\
	\textbf{Load} \\
    \textbf{Quantile}
  \end{tabular}}}}
  } &
\multirow{2}{*}{\rotatebox{90}{\parbox[b][0cm][b]{5cm}{\textbf{Naive}}}} &
\multirow{2}{*}{\rotatebox{90}{\parbox[b][0cm][b]{5cm}{\textbf{Lasso}}}} &
\multirow{2}{*}{\rotatebox{90}{\parbox[b][0cm][b]{5cm}{\textbf{XGBoost}}}} &
\multirow{2}{*}{\rotatebox{90}{\parbox[b][0cm][b]{5cm}{\textbf{NNet}}}} &
\multirow{2}{*}{\rotatebox{90}{\parbox[b][0cm][b]{5cm}{\textbf{Classic MO}}}} &
\multicolumn{6}{c|}{\textbf{Data-driven MO (i)}} \\
\cline{7-12}
& & & & & &
\rotatebox{90}{\parbox[b][0cm][b]{5cm}{\shortstack[b]{\textbf{(i) + Efficiencies (ii)}}}} &
\rotatebox{90}{\parbox[b][0cm][b]{5cm}{\textbf{(ii) + Cap Factors (iii)}}} &
\rotatebox{90}{\parbox[b][0cm][b]{5cm}{\textbf{(iii) + Net Import (iv)}}} &
\rotatebox{90}{\parbox[b][0cm][b]{5cm}{\textbf{(iv) + Tech Split (v)}}} &
\rotatebox{90}{\parbox[b][0cm][b]{5cm}{\textbf{(v) + Hydro (vi)}}} &
\rotatebox{90}{\parbox[b][0cm][b]{5cm}{\textbf{(vi) + Must Run (vii)}}} \\
\hline
  \hline
0\% - 5\% & \cellcolor[rgb]{1,0.6,0.6} $\text{40.13}$ & \cellcolor[rgb]{0.718,1,0.6} $\text{14.33}$ & \cellcolor[rgb]{0.696,1,0.6} $\text{13.51}$ & \cellcolor[rgb]{1,0.958,0.6} $\text{26.57}$ & \cellcolor[rgb]{0.717,1,0.6} $\text{14.31}$ & \cellcolor[rgb]{0.753,1,0.6} $\text{15.67}$ & \cellcolor[rgb]{0.715,1,0.6} $\text{14.23}$ & \cellcolor[rgb]{0.705,1,0.6} $\text{13.86}$ & \cellcolor[rgb]{0.6,1,0.6} $\textbf{\text{9.87}}$ & \cellcolor[rgb]{0.613,1,0.6} $\text{10.36}$ & \cellcolor[rgb]{0.864,1,0.6} $\text{19.84}$ \\ 
  5\% - 10\% & \cellcolor[rgb]{1,0.6,0.6} $\text{40.40}$ & \cellcolor[rgb]{0.739,1,0.6} $\text{14.79}$ & \cellcolor[rgb]{0.728,1,0.6} $\text{14.35}$ & \cellcolor[rgb]{1,0.984,0.6} $\text{25.53}$ & \cellcolor[rgb]{0.735,1,0.6} $\text{14.65}$ & \cellcolor[rgb]{0.693,1,0.6} $\text{13.03}$ & \cellcolor[rgb]{0.676,1,0.6} $\text{12.35}$ & \cellcolor[rgb]{0.657,1,0.6} $\text{11.60}$ & \cellcolor[rgb]{0.6,1,0.6} $\textbf{\text{9.41}}$ & \cellcolor[rgb]{0.635,1,0.6} $\text{10.77}$ & \cellcolor[rgb]{0.944,1,0.6} $\text{22.73}$ \\ 
  10\% - 15\% & \cellcolor[rgb]{1,0.6,0.6} $\text{34.80}$ & \cellcolor[rgb]{0.652,1,0.6} $\text{16.65}$ & \cellcolor[rgb]{0.71,1,0.6} $\text{18.07}$ & \cellcolor[rgb]{0.963,1,0.6} $\text{24.20}$ & \cellcolor[rgb]{0.692,1,0.6} $\text{17.63}$ & \cellcolor[rgb]{0.667,1,0.6} $\text{17.02}$ & \cellcolor[rgb]{0.765,1,0.6} $\text{19.39}$ & \cellcolor[rgb]{0.657,1,0.6} $\text{16.77}$ & \cellcolor[rgb]{0.726,1,0.6} $\text{18.44}$ & \cellcolor[rgb]{0.6,1,0.6} $\textbf{\text{15.39}}$ & \cellcolor[rgb]{1,0.959,0.6} $\text{26.09}$ \\ 
  15\% - 20\% & \cellcolor[rgb]{1,0.6,0.6} $\text{31.56}$ & \cellcolor[rgb]{0.753,1,0.6} $\text{18.62}$ & \cellcolor[rgb]{0.821,1,0.6} $\text{19.98}$ & \cellcolor[rgb]{0.995,1,0.6} $\text{23.45}$ & \cellcolor[rgb]{1,0.786,0.6} $\text{27.84}$ & \cellcolor[rgb]{0.945,1,0.6} $\text{22.45}$ & \cellcolor[rgb]{0.951,1,0.6} $\text{22.57}$ & \cellcolor[rgb]{0.798,1,0.6} $\text{19.53}$ & \cellcolor[rgb]{0.951,1,0.6} $\text{22.58}$ & \cellcolor[rgb]{0.6,1,0.6} $\textbf{\text{15.56}}$ & \cellcolor[rgb]{1,0.821,0.6} $\text{27.14}$ \\ 
  20\% - 25\% & \cellcolor[rgb]{1,0.6,0.6} $\text{29.75}$ & \cellcolor[rgb]{0.711,1,0.6} $\text{17.82}$ & \cellcolor[rgb]{0.733,1,0.6} $\text{18.21}$ & \cellcolor[rgb]{0.987,1,0.6} $\text{22.60}$ & \cellcolor[rgb]{1,0.623,0.6} $\text{29.36}$ & \cellcolor[rgb]{0.843,1,0.6} $\text{20.11}$ & \cellcolor[rgb]{1,1,0.6} $\text{22.82}$ & \cellcolor[rgb]{0.687,1,0.6} $\text{17.40}$ & \cellcolor[rgb]{0.886,1,0.6} $\text{20.85}$ & \cellcolor[rgb]{0.6,1,0.6} $\textbf{\text{15.90}}$ & \cellcolor[rgb]{0.76,1,0.6} $\text{18.67}$ \\ 
  25\% - 30\% & \cellcolor[rgb]{1,0.6,0.6} $\text{29.82}$ & \cellcolor[rgb]{0.664,1,0.6} $\text{16.38}$ & \cellcolor[rgb]{0.711,1,0.6} $\text{17.23}$ & \cellcolor[rgb]{0.868,1,0.6} $\text{20.11}$ & \cellcolor[rgb]{1,0.808,0.6} $\text{26.02}$ & \cellcolor[rgb]{0.861,1,0.6} $\text{19.98}$ & \cellcolor[rgb]{0.973,1,0.6} $\text{22.02}$ & \cellcolor[rgb]{0.634,1,0.6} $\text{15.83}$ & \cellcolor[rgb]{0.858,1,0.6} $\text{19.92}$ & \cellcolor[rgb]{0.6,1,0.6} $\textbf{\text{15.21}}$ & \cellcolor[rgb]{0.72,1,0.6} $\text{17.41}$ \\ 
  30\% - 35\% & \cellcolor[rgb]{1,0.6,0.6} $\text{26.26}$ & \cellcolor[rgb]{0.688,1,0.6} $\text{13.45}$ & \cellcolor[rgb]{0.77,1,0.6} $\text{14.93}$ & \cellcolor[rgb]{0.876,1,0.6} $\text{16.83}$ & \cellcolor[rgb]{0.974,1,0.6} $\text{18.60}$ & \cellcolor[rgb]{0.748,1,0.6} $\text{14.53}$ & \cellcolor[rgb]{0.825,1,0.6} $\text{15.91}$ & \cellcolor[rgb]{0.65,1,0.6} $\text{12.77}$ & \cellcolor[rgb]{0.736,1,0.6} $\text{14.32}$ & \cellcolor[rgb]{0.6,1,0.6} $\textbf{\text{11.87}}$ & \cellcolor[rgb]{0.627,1,0.6} $\text{12.35}$ \\ 
  35\% - 40\% & \cellcolor[rgb]{1,0.6,0.6} $\text{23.01}$ & \cellcolor[rgb]{0.646,1,0.6} $\text{11.60}$ & \cellcolor[rgb]{0.868,1,0.6} $\text{14.95}$ & \cellcolor[rgb]{0.848,1,0.6} $\text{14.66}$ & \cellcolor[rgb]{0.864,1,0.6} $\text{14.89}$ & \cellcolor[rgb]{0.71,1,0.6} $\text{12.56}$ & \cellcolor[rgb]{0.771,1,0.6} $\text{13.49}$ & \cellcolor[rgb]{0.661,1,0.6} $\text{11.82}$ & \cellcolor[rgb]{0.67,1,0.6} $\text{11.96}$ & \cellcolor[rgb]{0.6,1,0.6} $\textbf{\text{10.90}}$ & \cellcolor[rgb]{0.626,1,0.6} $\text{11.30}$ \\ 
  40\% - 45\% & \cellcolor[rgb]{1,0.6,0.6} $\text{22.28}$ & \cellcolor[rgb]{0.646,1,0.6} $\text{10.54}$ & \cellcolor[rgb]{0.731,1,0.6} $\text{11.86}$ & \cellcolor[rgb]{0.898,1,0.6} $\text{14.46}$ & \cellcolor[rgb]{0.787,1,0.6} $\text{12.73}$ & \cellcolor[rgb]{0.689,1,0.6} $\text{11.21}$ & \cellcolor[rgb]{0.723,1,0.6} $\text{11.73}$ & \cellcolor[rgb]{0.669,1,0.6} $\text{10.89}$ & \cellcolor[rgb]{0.643,1,0.6} $\text{10.49}$ & \cellcolor[rgb]{0.6,1,0.6} $\textbf{\text{9.82}}$ & \cellcolor[rgb]{0.619,1,0.6} $\text{10.12}$ \\ 
  45\% - 50\% & \cellcolor[rgb]{1,0.6,0.6} $\text{23.52}$ & \cellcolor[rgb]{0.668,1,0.6} $\text{10.12}$ & \cellcolor[rgb]{0.73,1,0.6} $\text{11.26}$ & \cellcolor[rgb]{0.946,1,0.6} $\text{15.21}$ & \cellcolor[rgb]{0.726,1,0.6} $\text{11.19}$ & \cellcolor[rgb]{0.69,1,0.6} $\text{10.52}$ & \cellcolor[rgb]{0.705,1,0.6} $\text{10.81}$ & \cellcolor[rgb]{0.667,1,0.6} $\text{10.10}$ & \cellcolor[rgb]{0.64,1,0.6} $\text{9.62}$ & \cellcolor[rgb]{0.6,1,0.6} $\textbf{\text{8.88}}$ & \cellcolor[rgb]{0.666,1,0.6} $\text{10.08}$ \\ 
  50\% - 55\% & \cellcolor[rgb]{1,0.6,0.6} $\text{20.98}$ & \cellcolor[rgb]{0.632,1,0.6} $\text{9.37}$ & \cellcolor[rgb]{0.668,1,0.6} $\text{9.92}$ & \cellcolor[rgb]{0.952,1,0.6} $\text{14.21}$ & \cellcolor[rgb]{0.708,1,0.6} $\text{10.52}$ & \cellcolor[rgb]{0.686,1,0.6} $\text{10.19}$ & \cellcolor[rgb]{0.686,1,0.6} $\text{10.19}$ & \cellcolor[rgb]{0.662,1,0.6} $\text{9.83}$ & \cellcolor[rgb]{0.623,1,0.6} $\text{9.24}$ & \cellcolor[rgb]{0.6,1,0.6} $\textbf{\text{8.89}}$ & \cellcolor[rgb]{0.691,1,0.6} $\text{10.26}$ \\ 
  55\% - 60\% & \cellcolor[rgb]{1,0.6,0.6} $\text{19.46}$ & \cellcolor[rgb]{0.611,1,0.6} $\text{7.96}$ & \cellcolor[rgb]{0.729,1,0.6} $\text{9.68}$ & \cellcolor[rgb]{0.97,1,0.6} $\text{13.20}$ & \cellcolor[rgb]{0.723,1,0.6} $\text{9.59}$ & \cellcolor[rgb]{0.678,1,0.6} $\text{8.93}$ & \cellcolor[rgb]{0.669,1,0.6} $\text{8.80}$ & \cellcolor[rgb]{0.64,1,0.6} $\text{8.39}$ & \cellcolor[rgb]{0.6,1,0.6} $\textbf{\text{7.80}}$ & \cellcolor[rgb]{0.608,1,0.6} $\text{7.91}$ & \cellcolor[rgb]{0.673,1,0.6} $\text{8.86}$ \\ 
  60\% - 65\% & \cellcolor[rgb]{1,0.6,0.6} $\text{18.88}$ & \cellcolor[rgb]{0.6,1,0.6} $\textbf{\text{7.63}}$ & \cellcolor[rgb]{0.771,1,0.6} $\text{10.03}$ & \cellcolor[rgb]{0.938,1,0.6} $\text{12.39}$ & \cellcolor[rgb]{0.775,1,0.6} $\text{10.09}$ & \cellcolor[rgb]{0.782,1,0.6} $\text{10.19}$ & \cellcolor[rgb]{0.754,1,0.6} $\text{9.79}$ & \cellcolor[rgb]{0.653,1,0.6} $\text{8.37}$ & \cellcolor[rgb]{0.668,1,0.6} $\text{8.59}$ & \cellcolor[rgb]{0.64,1,0.6} $\text{8.19}$ & \cellcolor[rgb]{0.724,1,0.6} $\text{9.38}$ \\ 
  65\% - 70\% & \cellcolor[rgb]{1,0.6,0.6} $\text{18.65}$ & \cellcolor[rgb]{0.609,1,0.6} $\text{7.89}$ & \cellcolor[rgb]{0.752,1,0.6} $\text{9.84}$ & \cellcolor[rgb]{0.949,1,0.6} $\text{12.52}$ & \cellcolor[rgb]{0.732,1,0.6} $\text{9.56}$ & \cellcolor[rgb]{0.747,1,0.6} $\text{9.77}$ & \cellcolor[rgb]{0.704,1,0.6} $\text{9.19}$ & \cellcolor[rgb]{0.607,1,0.6} $\text{7.87}$ & \cellcolor[rgb]{0.623,1,0.6} $\text{8.08}$ & \cellcolor[rgb]{0.6,1,0.6} $\textbf{\text{7.77}}$ & \cellcolor[rgb]{0.663,1,0.6} $\text{8.63}$ \\ 
  70\% - 75\% & \cellcolor[rgb]{1,0.6,0.6} $\text{18.59}$ & \cellcolor[rgb]{0.6,1,0.6} $\textbf{\text{8.44}}$ & \cellcolor[rgb]{0.757,1,0.6} $\text{10.43}$ & \cellcolor[rgb]{0.975,1,0.6} $\text{13.20}$ & \cellcolor[rgb]{0.771,1,0.6} $\text{10.61}$ & \cellcolor[rgb]{0.766,1,0.6} $\text{10.55}$ & \cellcolor[rgb]{0.729,1,0.6} $\text{10.08}$ & \cellcolor[rgb]{0.641,1,0.6} $\text{8.96}$ & \cellcolor[rgb]{0.663,1,0.6} $\text{9.24}$ & \cellcolor[rgb]{0.623,1,0.6} $\text{8.73}$ & \cellcolor[rgb]{0.691,1,0.6} $\text{9.60}$ \\ 
  75\% - 80\% & \cellcolor[rgb]{1,0.6,0.6} $\text{16.84}$ & \cellcolor[rgb]{0.713,1,0.6} $\text{9.90}$ & \cellcolor[rgb]{0.735,1,0.6} $\text{10.12}$ & \cellcolor[rgb]{1,0.925,0.6} $\text{13.56}$ & \cellcolor[rgb]{0.948,1,0.6} $\text{12.27}$ & \cellcolor[rgb]{0.777,1,0.6} $\text{10.55}$ & \cellcolor[rgb]{0.707,1,0.6} $\text{9.84}$ & \cellcolor[rgb]{0.602,1,0.6} $\text{8.78}$ & \cellcolor[rgb]{0.633,1,0.6} $\text{9.09}$ & \cellcolor[rgb]{0.6,1,0.6} $\textbf{\text{8.76}}$ & \cellcolor[rgb]{0.809,1,0.6} $\text{10.87}$ \\ 
  80\% - 85\% & \cellcolor[rgb]{1,0.6,0.6} $\text{17.56}$ & \cellcolor[rgb]{0.742,1,0.6} $\text{11.65}$ & \cellcolor[rgb]{0.735,1,0.6} $\text{11.59}$ & \cellcolor[rgb]{1,0.821,0.6} $\text{15.58}$ & \cellcolor[rgb]{1,0.913,0.6} $\text{14.75}$ & \cellcolor[rgb]{0.764,1,0.6} $\text{11.85}$ & \cellcolor[rgb]{0.721,1,0.6} $\text{11.47}$ & \cellcolor[rgb]{0.6,1,0.6} $\textbf{\text{10.38}}$ & \cellcolor[rgb]{0.641,1,0.6} $\text{10.75}$ & \cellcolor[rgb]{0.632,1,0.6} $\text{10.67}$ & \cellcolor[rgb]{0.804,1,0.6} $\text{12.21}$ \\ 
  85\% - 90\% & \cellcolor[rgb]{1,0.6,0.6} $\text{18.71}$ & \cellcolor[rgb]{0.649,1,0.6} $\text{12.31}$ & \cellcolor[rgb]{0.771,1,0.6} $\text{13.35}$ & \cellcolor[rgb]{1,0.906,0.6} $\text{16.10}$ & \cellcolor[rgb]{1,0.667,0.6} $\text{18.14}$ & \cellcolor[rgb]{0.751,1,0.6} $\text{13.18}$ & \cellcolor[rgb]{0.782,1,0.6} $\text{13.44}$ & \cellcolor[rgb]{0.601,1,0.6} $\text{11.90}$ & \cellcolor[rgb]{0.6,1,0.6} $\textbf{\text{11.89}}$ & \cellcolor[rgb]{0.667,1,0.6} $\text{12.46}$ & \cellcolor[rgb]{0.791,1,0.6} $\text{13.52}$ \\ 
  90\% - 95\% & \cellcolor[rgb]{1,0.871,0.6} $\text{22.06}$ & \cellcolor[rgb]{0.881,1,0.6} $\text{19.01}$ & \cellcolor[rgb]{0.83,1,0.6} $\text{18.39}$ & \cellcolor[rgb]{1,0.6,0.6} $\text{25.39}$ & \cellcolor[rgb]{0.939,1,0.6} $\text{19.73}$ & \cellcolor[rgb]{0.672,1,0.6} $\text{16.44}$ & \cellcolor[rgb]{0.812,1,0.6} $\text{18.16}$ & \cellcolor[rgb]{0.653,1,0.6} $\text{16.21}$ & \cellcolor[rgb]{0.6,1,0.6} $\textbf{\text{15.56}}$ & \cellcolor[rgb]{0.691,1,0.6} $\text{16.68}$ & \cellcolor[rgb]{0.723,1,0.6} $\text{17.07}$ \\ 
  95\% - 100\% & \cellcolor[rgb]{0.75,1,0.6} $\text{25.92}$ & \cellcolor[rgb]{0.707,1,0.6} $\text{23.90}$ & \cellcolor[rgb]{0.66,1,0.6} $\text{21.72}$ & \cellcolor[rgb]{0.87,1,0.6} $\text{31.51}$ & \cellcolor[rgb]{0.726,1,0.6} $\text{24.79}$ & \cellcolor[rgb]{0.6,1,0.6} $\textbf{\text{18.94}}$ & \cellcolor[rgb]{1,0.6,0.6} $\text{56.16}$ & \cellcolor[rgb]{0.646,1,0.6} $\text{21.08}$ & \cellcolor[rgb]{0.618,1,0.6} $\text{19.77}$ & \cellcolor[rgb]{0.662,1,0.6} $\text{21.83}$ & \cellcolor[rgb]{0.645,1,0.6} $\text{21.03}$ \\ 
   \hline
   \hline
Price < 0 & \cellcolor[rgb]{1,0.6,0.6} $\text{39.40}$ & \cellcolor[rgb]{0.794,1,0.6} $\text{16.80}$ & \cellcolor[rgb]{0.755,1,0.6} $\text{15.33}$ & \cellcolor[rgb]{1,0.858,0.6} $\text{29.76}$ & \cellcolor[rgb]{0.703,1,0.6} $\text{13.40}$ & \cellcolor[rgb]{0.787,1,0.6} $\text{16.54}$ & \cellcolor[rgb]{0.676,1,0.6} $\text{12.41}$ & \cellcolor[rgb]{0.697,1,0.6} $\text{13.16}$ & \cellcolor[rgb]{0.657,1,0.6} $\text{11.70}$ & \cellcolor[rgb]{0.6,1,0.6} $\textbf{\text{9.56}}$ & \cellcolor[rgb]{0.824,1,0.6} $\text{17.92}$ \\ 
   \hline
\end{tabular}}}
\caption{Test MAE (EUR/MWh) for models of the optimal path in \ref{fig:forward_tree} broken down by quantile bins of residual load and for negative prices.}
\label{tab:errors_by_res}
\end{table}

\clearpage
\section{Discussion}
\label{sec:discussion}

\subsection{Model comparison and interpretability}
\begin{wrapfigure}[16]{r}{8cm}
	\includegraphics[width=8cm]{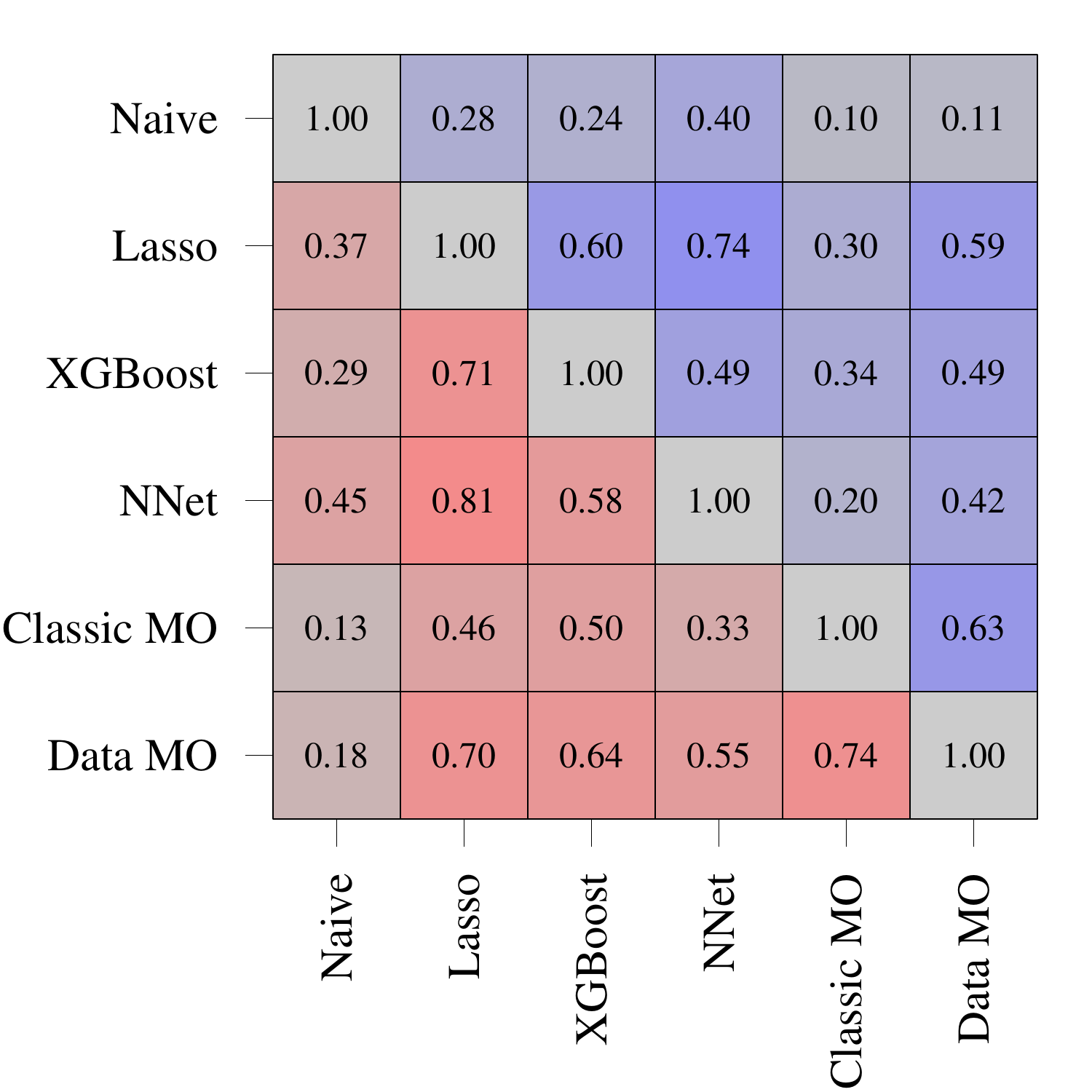}
	\caption{Correlation of forecast errors: Pearson's correlation on the lower triangle, distance correlation on the upper triangle.}
	\label{fig:result_correl}
\end{wrapfigure}

The superior performance of the \textbf{data-driven merit order} model over established machine-learning benchmarks, despite not using autoregressive inputs, is noteworthy.
Figure \ref{fig:result_correl} compares the models in terms of linear and non-linear similarities as measured by the Pearson correlation and distance correlation coefficients respectively. The data-driven merit order model is most similar to the classical merit order, confirming its fundamental nature, yet it also shows considerable similarity to \textbf{Lasso} and \textbf{XGBoost}. Nonetheless, results in Figure \ref{fig:result_mo_optim}.b and Table \ref{tab:errors_by_res} indicate that each model captures distinct aspects, suggesting potential for model combination.

A key strength of the proposed model lies in its ability to estimate market and plant-level parameters (see Appendix Table \ref{tab:theta_bounds}). Most estimates align with expert values, though coal and oil efficiencies tend to fall at their lower bounds. Capacity factors are often estimated above 1, implying underreported capacities at the unit level. In contrast, must-run shares are less conclusive as coal and gas plants show unexpectedly low values. This highlights the trade-off in using an unconstrained optimization for price prediction. Future work could impose tighter constraints or adopt a joint objective function for both price and production, exploiting the customizable loss function, thus making estimated parameters more useful for further analysis.

\subsection{Structural insights from the merit order model}
\begin{figure}[hbtp]
    \subfloat[Marginal costs with marginal technology mix (2024)]{
        \includegraphics[width=0.9\textwidth]{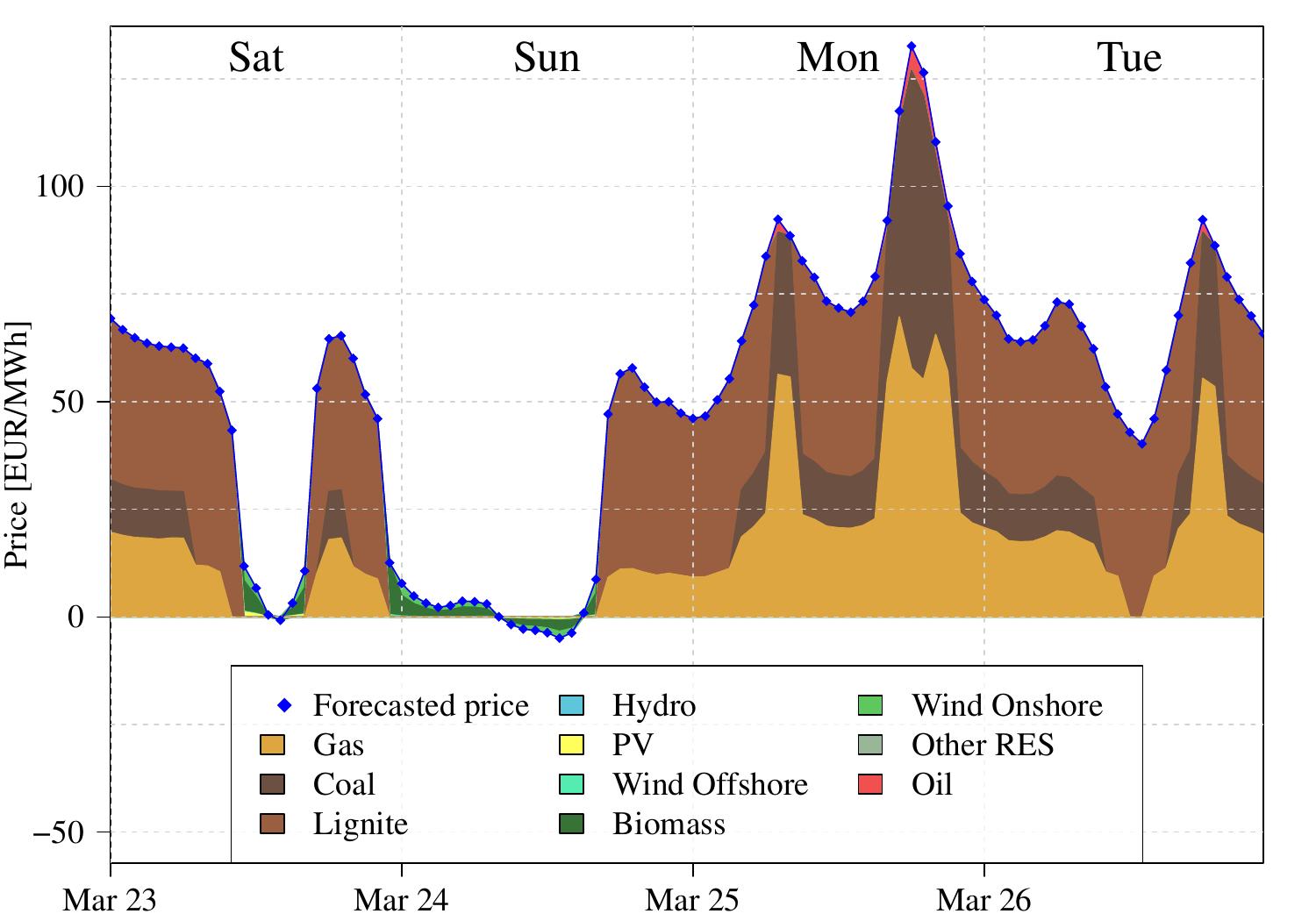}}
	\linebreak
    \subfloat[Fuel switches]{
        \includegraphics[width=0.9\textwidth]{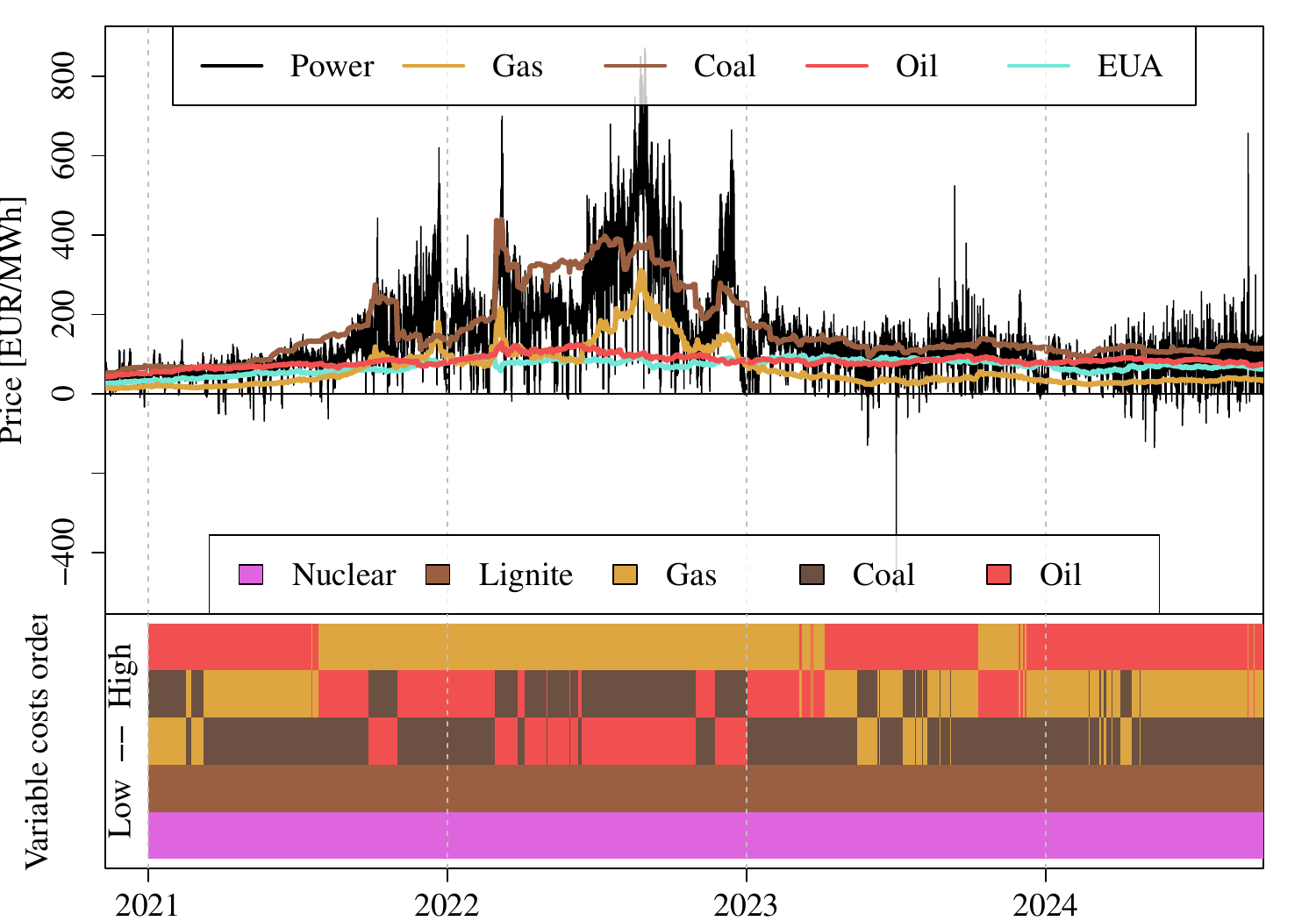}}
	\caption{Information extracted from the best data-driven merit order model.}
	\label{fig:result_info_extract}
\end{figure}
Beyond forecasting, the merit order model provides deeper structural insights. Unlike econometric models, which model prices directly and implicitly weigh the influence of all plant types, the merit order framework focuses on the market-clearing point where supply meets demand and characterizes the price-setting dynamics more explicitly.

The clearing price corresponds to the marginal cost of the last dispatched (at-the-money, ATM) plant. Generators with lower costs are in-the-money (ITM) and are dispatched; those with higher costs are out-of-the-money (OTM) and idle. This setup enables the identification of price-setting technologies and their marginal influence (see Figure \ref{fig:result_info_extract}.a).

The model also reveals fuel-switching dynamics (see Figure \ref{fig:result_info_extract}.b), which occur when relative cost changes alter the dispatch order. For example, during the European energy crisis in mid-2021, surging gas prices triggered a switch from oil to gas. This trend reversed at the beginning of 2023 as gas prices normalized.
By late 2024, fuel prices had stabilized. According to the data-driven merit order model, oil remained the most expensive fuel, followed by gas, coal, and lignite. Nuclear and lignite continued to serve as base-load technologies due to their low variable costs and were largely unaffected by fuel switches.

\section{Summary and Conclusion}
\label{sec:conclusion}

We propose a novel \textbf{data-driven merit order model} that combines the interpretability of fundamental models with the flexibility of data-driven approaches. It directly embeds the classical merit order, meaning that if the expert-defined fundamental model were optimal, the algorithm would select it. Thus, the classical model becomes a nested special case within our framework.
By optimizing key parameters using historical data, the proposed model significantly improves price forecasting accuracy—outperforming state-of-the-art methods like Lasso, XGBoost, and neural networks, even without autoregressive features.
We introduce several impactful extensions to the merit order model, including capacity corrections for underreportings, hydro power modeling, imports/exports modeling, technology splits (e.g., gas split into CCGT vs. OCGT), and must-run shares.These embedded extensions enable the model to better capture physical system behavior and price-setting mechanisms.
Though now calibrated to data, the model remains a fundamental model in essence, retaining the ability to explain causal drivers such as marginal plant types, dispached plants, production and fuel switches, which are inaccessible to traditional machine learning models. Its structure allows for clear interpretation and inspection of economic effects.

To our knowledge, this is the first full-fledged, intrinsic integration of fundamental and data-driven modeling in electricity price forecasting. The results are highly promising and open up a new frontier for research and application in the field.
The model provides a solid foundation for future extensions, including a more accurate reflection of policy constraints (e.g., RES subsidies), more refined modeling of cross-border flows, probabilistic forecasting, the inclusion of storage and market coupling effects, extension of the forecasting horizon and smoothening of the merit order curve.

\clearpage
\section*{Declaration of Generative AI and AI-assisted technologies}
During the preparation of this work the authors used GitHub Copilot and OpenAI ChatGPT-3.5 in order to generate rephrasing suggestions to improve the language and readability of this paper. After using these services, the authors reviewed and edited the content as needed and take full responsibility for the content of the published article.

\section*{Acknowledgements}
Funding: This work was supported by the German Research Foundation (DFG, Deutsche Forschungsgemeinschaft) project numbers 505565850 and 520388526 to P.G. and F.Z.

\section*{Data Availability}
The raw data required to reproduce the findings in this work include the freely available resources: day-ahead power prices, power plant capacities, generation and unavailabilities, power imports and exports, actual values and day-ahead forecasts for load, solar, onshore wind, and offshore wind infeed. This data can be downloaded from the ENTSOE Transparency Platform (https://transparency.entsoe.eu/).
Data not freely available include daily closing prices for futures of gas, coal, oil, and European Emission Allowances (EUA), accessible via the information platform Refinitiv Eikon at a cost (https://eikon.refinitiv.com/).

\clearpage
\section{Appendix}

\subsection{Hydro power forecast model}
 \begin{align}
	\label{eq:hydro}
	\text{Hydro}_{d,h} &= \beta_0 + \underbrace{\sum_{s=1}^{11} \beta_{\text{cross},s} \text{Hydro}_{d-1,s} + \sum_{s=12}^{24} \beta_{\text{cross},s} \text{Hydro}_{d-2,s}}_{\text{Cross-hour lags}} + \underbrace{\sum_{l=2}^{14} \beta_{\text{lag},l} \text{Hydro}_{d-l,h}}_{\text{Same hour lags}}  \\
					   &+ \underbrace{\sum_{i=1}^{4} \beta_{\text{SoY},i} \text{Season}^{(i)}(d,h)}_{\text{Annual seasonality}} + \underbrace{\sum_{i=1}^{7} \beta_{\text{DoW},i} \text{WeekDay}^{(i)}_{d}}_{\text{Weekly Seasonality}}  + \varepsilon_t \quad \forall \quad h \in \{ 0, \dots, 23 \} \nonumber
\end{align}
where $d$ is the day and $h$ the hour. The model is estimated using LASSO and the optimal lambda parameter is chosen by the BIC (Bayesian information criterion).

 \subsection{Net import forecast model}
 \begin{align}
	\label{eq:net_import}
	\text{NetImport}_{d,h}^{\text{bznHome}} &= \beta_0 + \underbrace{\sum_{\text{bzn} \in \mathcal{B}} \beta_{\text{Load}}^{\text{bzn}} \text{LoadDA}_{d,h}^{\text{bzn}}}_{\text{Day-ahead load}} 
	+ \underbrace{\sum_{\text{bzn} \in \mathcal{B}} \sum_{pl \in \mathcal{R}} \beta_{\text{pl, RES}}^{\text{bzn}} \text{RESDA}_{pl,d,h}}_{\text{Renewables day-ahead}} \\
	&+ \underbrace{\sum_{\text{bzn} \in \mathcal{B}} \sum_{pl \in \mathcal{C}} \beta_{\text{pl, gen}}^{\text{bzn}} \text{Generation}_{pl,(d-1) \textbf{1}_{h \in \{1, \dots, 11 \}} + (d-2) \textbf{1}_{h \in \{12, \dots, 24 \}}}}_{\text{Generation of non-RES}} \nonumber \\
	&+ \underbrace{\sum_{\text{i=1}}^{14} \beta_{\text{lag},i} \text{NetImport}_{d-i,(d-1) \textbf{1}_{h \in \{1, \dots, 11 \}} + (d-2) \textbf{1}_{h \in \{12, \dots, 24 \}}}^{\text{bznHome}}}_{\text{Lagged net imports}} \nonumber \\ 
	&+ \varepsilon_t \quad \forall \quad h \in \{ 0, \dots, 23 \} \nonumber
\end{align}
where $d$ is the day, is $h$ the hour, $\mathcal{B}$ is the set of all bidding zones of physical flows to and from and including the modelled zone $\text{bznHome}$, $\mathcal{R}$ is the set of all renewable power plants, $\mathcal{C}$ is the set of all conventional power plants, and $\textbf{1}$ is the indicator function.

\subsection{Naive model}
\begin{align}
	\label{eq:naive}
	\text{Price}_{d,h} =  
	\begin{cases}
		\text{Price}_{d-7,h} & \text{if } \text{WeekDay}(d) \in \{ \text{Mon}, \text{Sat}, \text{Sun} \}, \\
		\text{Price}_{d-1,h} & \text{otherwise},
	\end{cases}
\end{align}
where $d$ is the day and $h$ the hour.

\subsection{Expert model}
\begin{align}
	\label{eq:expert}
	\text{Price}_{d,h} &=  \beta_0 + \underbrace{\sum_{i=1}^{14} \beta_{\text{lag},i} \text{Price}_{d-i,h}}_{\text{Autoregressive effects}} + \underbrace{\beta_{\text{max}} \max(\text{Price}_{d-1,\cdot}) + \beta_{\text{min}} \min(\text{Price}_{d-1,\cdot})}_{\text{Non-linear effects: min and max of yesterday}} \\
	& + \underbrace{\beta_{\text{last}} \text{Price}_{d-1,23}}_{\text{Cross-period effects}} + \underbrace{\beta_{\text{load}} \text{LoadDA}_{d,h}}_{\text{Load day-ahead forecast}} + \underbrace{\beta_{\text{res}} \text{RESDA}_{d,h}}_{\text{RES day-ahead forecast}}  \nonumber \\
	& + \underbrace{\beta_{\text{gas}} \text{Gas}_{d-2} + \beta_{\text{coal}} \text{Coal}_{d-2} + \beta_{\text{oil}} \text{Oil}_{d-2} + \beta_{\text{CO}_2} \text{CO}_{2,d-2}}_{\text{Fuel prices}} \nonumber \\
	& + \underbrace{\sum_{i=1}^{7} \beta_{\text{DoW},i} \text{WeekDay}^{(i)}(d)}_{\text{Weekly seasonality}} + \underbrace{\sum_{i=1}^{4} \beta_{\text{SoY,i}} \text{Season}^{(i)}(d,h)}_{\text{Annual seasonality}}  + \varepsilon_t \quad \forall \quad h \in \{ 0, \dots, 23 \} \nonumber
\end{align}
where $d$ is the day and $h$ the hour.

\subsection{Parameter values}
\begin{table}[ht]
	\setlength{\tabcolsep}{1pt}
	\setlength\extrarowheight{-4pt}	
\centering
\scalebox{0.65}{
\begin{tabular}{|p{2.2cm}|R{2cm}|R{2cm}|R{2.5cm}|R{2.5cm}|R{2.3cm}|R{2.3cm}|R{2.3cm}|R{2.3cm}|R{2.3cm}|}
	\hline
\makecell[c]{\textbf{Parameter}} & 
\makecell[c]{\textbf{Lower}\\\textbf{($\Theta^\text{L}$)}} & 
\makecell[c]{\textbf{Upper}\\\textbf{($\Theta^\text{U}$)}} & 
\makecell[c]{\textbf{Classic MO}\\\textbf{($\Theta_\text{init}$)}} & 
\makecell[c]{\textbf{Data MO}\\\textbf{+}\\\textbf{Efficiencies}} & 
\makecell[c]{\\ \textbf{+}\\\textbf{Cap Factor}} & 
\makecell[c]{\\ \textbf{+}\\\textbf{Net Import}} & 
\makecell[c]{\\ \textbf{+}\\\textbf{Tech Split}} & 
\makecell[c]{\\ \textbf{+}\\\textbf{Hydro}} & 
\makecell[c]{\\ \textbf{+}\\\textbf{Must Run}} \\
	\hline
 $\eta^{\text{L}}_\text{gas}$ & 0.10 & 0.50 & 0.25 & 0.40 & 0.42 & 0.29 & 0.33 & 0.23 & 0.33 \\ 
  $\eta^{\text{L}}_\text{coal}$ & 0.10 & 0.50 & 0.35 & 0.12 & 0.14 & 0.26 & 0.13 & 0.17 & 0.11 \\ 
  $\eta^{\text{L}}_\text{lignite}$ & 0.10 & 0.50 & 0.30 & 0.22 & 0.25 & 0.44 & 0.31 & 0.33 & 0.21 \\ 
  $\eta^{\text{L}}_\text{oil}$ & 0.10 & 0.50 & 0.24 & 0.20 & 0.15 & 0.33 & 0.32 & 0.45 & 0.11 \\ 
  $\eta^{\text{L}}_\text{nuclear}$ & 0.10 & 0.50 & 0.32 & 0.37 & 0.25 & 0.46 & 0.19 & 0.47 & 0.10 \\ 
  $\eta^{\text{L}}_\text{gas2}$ & 0.10 & 0.50 & 0.10 & 0.44 & 0.40 & 0.37 & 0.32 & 0.24 & 0.47 \\ 
   \hline
$\eta^{\text{U}}_\text{gas}$ & 0.10 & 1.00 & 0.40 & 0.71 & 0.68 & 0.72 & 0.54 & 0.64 & 0.42 \\ 
  $\eta^{\text{U}}_\text{coal}$ & 0.10 & 1.00 & 0.46 & 0.20 & 0.64 & 0.49 & 0.51 & 0.52 & 0.57 \\ 
  $\eta^{\text{U}}_\text{lignite}$ & 0.10 & 1.00 & 0.43 & 0.69 & 0.63 & 0.80 & 0.72 & 0.68 & 0.52 \\ 
  $\eta^{\text{U}}_\text{oil}$ & 0.10 & 1.00 & 0.44 & 0.36 & 0.16 & 0.65 & 0.74 & 0.78 & 0.19 \\ 
  $\eta^{\text{U}}_\text{nuclear}$ & 0.10 & 1.00 & 0.42 & 0.43 & 0.58 & 0.51 & 0.31 & 0.83 & 0.50 \\ 
  $\eta^{\text{U}}_\text{gas2}$ & 0.10 & 1.00 & 0.20 & 0.78 & 0.56 & 0.68 & 0.49 & 0.62 & 0.53 \\ 
   \hline
$b^\text{L}_{\text{resHydro}}$ & -500.00 & 0.00 & -500.00 & -175.25 & -456.42 & -422.35 & -51.41 & -464.70 & -28.07 \\ 
  $b^\text{L}_{\text{resPV}}$ & -500.00 & 0.00 & -500.00 & -468.87 & -474.47 & -224.24 & -494.73 & -364.57 & -488.55 \\ 
  $b^\text{L}_{\text{resWindOn}}$ & -500.00 & 0.00 & -70.00 & -1.45 & -416.73 & -180.61 & -159.54 & -121.73 & -110.47 \\ 
  $b^\text{L}_{\text{resWindOff}}$ & -500.00 & 0.00 & -150.00 & -38.83 & -203.47 & -75.37 & -385.07 & -440.78 & -13.28 \\ 
  $b^\text{L}_{\text{resBiomass}}$ & -500.00 & 0.00 & -200.00 & -95.90 & -12.56 & -201.79 & -2.76 & -24.87 & -262.39 \\ 
  $b^\text{L}_{\text{resOther}}$ & -500.00 & 0.00 & -500.00 & -497.95 & -142.49 & -177.27 & -21.66 & -103.21 & -476.19 \\ 
   \hline
$b^\text{U}_{\text{resHydro}}$ & 0.00 & 20.00 & 20.00 & 12.67 & 7.52 & 19.04 & 17.89 & 17.91 & 17.68 \\ 
  $b^\text{U}_{\text{resPV}}$ & 0.00 & 20.00 & 20.00 & 13.71 & 9.41 & 12.93 & 4.51 & 17.64 & 19.48 \\ 
  $b^\text{U}_{\text{resWindOn}}$ & 0.00 & 20.00 & 20.00 & 18.14 & 18.90 & 14.31 & 15.07 & 11.89 & 9.53 \\ 
  $b^\text{U}_{\text{resWindOff}}$ & 0.00 & 20.00 & 20.00 & 11.95 & 11.79 & 4.68 & 0.01 & 12.91 & 0.40 \\ 
  $b^\text{U}_{\text{resBiomass}}$ & 0.00 & 20.00 & 20.00 & 15.07 & 9.63 & 3.36 & 1.79 & 13.93 & 4.68 \\ 
  $b^\text{U}_{\text{resOther}}$ & 0.00 & 20.00 & 20.00 & 4.70 & 2.09 & 14.98 & 3.97 & 2.44 & 18.12 \\ 
   \hline
$\text{cf}_{\text{gas}}$ & 1.00 & 2.00 & 1.50 & 1.50 & 1.66 & 1.62 & 1.63 & 1.34 & 1.52 \\ 
  $\text{cf}_\text{coal}$ & 1.00 & 2.00 & 1.50 & 1.50 & 1.23 & 1.13 & 1.95 & 1.28 & 1.70 \\ 
  $\text{cf}_\text{lignite}$ & 1.00 & 2.00 & 1.50 & 1.50 & 1.11 & 1.03 & 1.25 & 1.04 & 1.35 \\ 
  $\text{cf}_\text{oil}$ & 1.00 & 2.00 & 1.50 & 1.50 & 1.75 & 1.43 & 1.84 & 1.76 & 1.86 \\ 
  $\text{cf}_\text{resHydro}$ & 1.00 & 2.00 & 0.00 & 0.00 & 0.00 & 0.00 & 0.00 & 1.10 & 1.53 \\ 
  $\text{cf}_\text{netImport}$ & 0.00 & 1.00 & 0.00 & 0.00 & 0.00 & 0.54 & 0.39 & 0.55 & 0.50 \\ 
  $\text{cf}_\text{resBiomass}$ & 1.00 & 2.00 & 1.50 & 1.50 & 1.35 & 1.20 & 1.16 & 1.45 & 1.01 \\ 
  $\text{cf}_\text{gas2}$ & 1.00 & 2.00 & 1.50 & 1.50 & 1.96 & 1.71 & 1.04 & 1.08 & 1.27 \\ 
   \hline
$\text{mr}_{\text{gas}}$ & 0.00 & 1.00 & 0.00 & 0.00 & 0.00 & 0.00 & 0.00 & 0.00 & 0.28 \\ 
  $\text{mr}_\text{coal}$ & 0.00 & 1.00 & 0.00 & 0.00 & 0.00 & 0.00 & 0.00 & 0.00 & 0.06 \\ 
  $\text{mr}_\text{lignite}$ & 0.00 & 1.00 & 0.00 & 0.00 & 0.00 & 0.00 & 0.00 & 0.00 & 0.02 \\ 
  $\text{mr}_\text{oil}$ & 0.00 & 1.00 & 0.00 & 0.00 & 0.00 & 0.00 & 0.00 & 0.00 & 0.43 \\ 
  $\text{mr}_\text{nuclear}$ & 0.00 & 1.00 & 0.00 & 0.00 & 0.00 & 0.00 & 0.00 & 0.00 & 0.54 \\ 
  $\text{mr}_\text{resHydro}$ & 0.00 & 1.00 & 0.00 & 0.00 & 0.00 & 0.00 & 0.00 & 0.00 & 0.73 \\ 
  $\text{mr}_\text{resBiomass}$ & 0.00 & 1.00 & 0.00 & 0.00 & 0.00 & 0.00 & 0.00 & 0.00 & 0.12 \\ 
  $\text{mr}_\text{gas2}$ & 0.00 & 1.00 & 0.00 & 0.00 & 0.00 & 0.00 & 0.00 & 0.00 & 0.10 \\ 
   \hline
$\text{gs}$ & 0.00 & 1.00 & 0.00 & 0.00 & 0.00 & 0.00 & 0.58 & 0.41 & 0.50 \\ 
  \hline
  \end{tabular}
  }
	\caption{Parameter bounds, initial values, and optimized values for $\Theta_\text{ext}$ across various model setups, following the optimal forward selection path in Figure \ref{fig:forward_tree}. Each column corresponds to a different model setup and optimization, with one variable added at each step. The right-most column represents the full model, where all six parameter types are optimized simultaneously.}
	\label{tab:theta_bounds}
\end{table}

\clearpage
\vspace{-5mm} 
 \bibliographystyle{unsrtnat1.bst}

\bibliography{bibliography}

\begin{thebibliography}{65}
\providecommand{\natexlab}[1]{#1}
\providecommand{\url}[1]{\texttt{#1}}
\expandafter\ifx\csname urlstyle\endcsname\relax
  \providecommand{\doi}[1]{doi: #1}\else
  \providecommand{\doi}{doi: \begingroup \urlstyle{rm}\Url}\fi

\bibitem[Petropoulos et~al.(2022)Petropoulos, Apiletti, Assimakopoulos, Babai, Barrow, Taieb, Bergmeir, Bessa, Bijak, Boylan, et~al.]{petropoulos2022forecasting}
F.~Petropoulos, D.~Apiletti, V.~Assimakopoulos, M.~Z. Babai, D.~K. Barrow, S.~B. Taieb, et~al.
\newblock Forecasting: theory and practice.
\newblock \emph{International Journal of Forecasting}, 38\penalty0 (3):\penalty0 705--871, 2022.

\bibitem[Weron(2014)]{weron2014electricity}
R.~Weron.
\newblock Electricity price forecasting: A review of the state-of-the-art with a look into the future.
\newblock \emph{International journal of forecasting}, 30\penalty0 (4):\penalty0 1030--1081, 2014.

\bibitem[Ziel and Steinert(2016)]{ziel2016electricity}
F.~Ziel and R.~Steinert.
\newblock Electricity price forecasting using sale and purchase curves: The x-model.
\newblock \emph{Energy Economics}, 59:\penalty0 435--454, 2016.

\bibitem[Hastie et~al.(2009)Hastie, Tibshirani, Friedman, and Friedman]{hastie2009elements}
T.~Hastie, R.~Tibshirani, J.~H. Friedman, and J.~H. Friedman.
\newblock \emph{The elements of statistical learning: data mining, inference, and prediction}, volume~2.
\newblock Springer, 2009.

\bibitem[L{\"u}tkepohl(2005)]{lutkepohl2005new}
H.~L{\"u}tkepohl.
\newblock \emph{New introduction to multiple time series analysis}.
\newblock Springer Science \& Business Media, 2005.

\bibitem[Brockwell and Davis(2002)]{brockwell2002introduction}
P.~J. Brockwell and R.~A. Davis.
\newblock \emph{Introduction to time series and forecasting}.
\newblock Springer, 2002.

\bibitem[Cuaresma et~al.(2004)Cuaresma, Hlouskova, Kossmeier, and Obersteiner]{cuaresma2004forecasting}
J.~C. Cuaresma, J.~Hlouskova, S.~Kossmeier, and M.~Obersteiner.
\newblock Forecasting electricity spot-prices using linear univariate time-series models.
\newblock \emph{Applied Energy}, 77\penalty0 (1):\penalty0 87--106, 2004.

\bibitem[Narajewski and Ziel(2020)]{narajewski2020econometric}
M.~Narajewski and F.~Ziel.
\newblock Econometric modelling and forecasting of intraday electricity prices.
\newblock \emph{Journal of Commodity Markets}, 19:\penalty0 100107, 2020.

\bibitem[Uniejewski et~al.(2019)Uniejewski, Marcjasz, and Weron]{uniejewski2019understanding}
B.~Uniejewski, G.~Marcjasz, and R.~Weron.
\newblock Understanding intraday electricity markets: Variable selection and very short-term price forecasting using lasso.
\newblock \emph{International Journal of Forecasting}, 35\penalty0 (4):\penalty0 1533--1547, 2019.

\bibitem[Wintenberger(2017)]{wintenberger2017optimal}
O.~Wintenberger.
\newblock Optimal learning with bernstein online aggregation.
\newblock \emph{Machine Learning}, 106:\penalty0 119--141, 2017.

\bibitem[Cesa-Bianchi and Orabona(2021)]{cesa2021online}
N.~Cesa-Bianchi and F.~Orabona.
\newblock Online learning algorithms.
\newblock \emph{Annual review of statistics and its application}, 8\penalty0 (1):\penalty0 165--190, 2021.

\bibitem[Adjakossa et~al.(2024)Adjakossa, Goude, and Wintenberger]{adjakossa2024kalman}
E.~Adjakossa, Y.~Goude, and O.~Wintenberger.
\newblock Kalman recursions aggregated online.
\newblock \emph{Statistical Papers}, 65\penalty0 (2):\penalty0 909--944, 2024.

\bibitem[Ziel(2022)]{ziel2022smoothed}
F.~Ziel.
\newblock Smoothed bernstein online aggregation for short-term load forecasting in ieee dataport competition on day-ahead electricity demand forecasting: Post-covid paradigm.
\newblock \emph{IEEE Open Access Journal of Power and Energy}, 9:\penalty0 202--212, 2022.

\bibitem[Ringkj{\o}b et~al.(2018)Ringkj{\o}b, Haugan, and Solbrekke]{ringkjob2018review}
H.-K. Ringkj{\o}b, P.~M. Haugan, and I.~M. Solbrekke.
\newblock A review of modelling tools for energy and electricity systems with large shares of variable renewables.
\newblock \emph{Renewable and Sustainable Energy Reviews}, 96:\penalty0 440--459, 2018.

\bibitem[Brown et~al.(2017)Brown, H{\"o}rsch, and Schlachtberger]{brown2017pypsa}
T.~Brown, J.~H{\"o}rsch, and D.~Schlachtberger.
\newblock Pypsa: Python for power system analysis.
\newblock \emph{arXiv preprint arXiv:1707.09913}, 2017.

\bibitem[Beran et~al.(2021)Beran, Vogler, and Weber]{beran2021multi}
P.~Beran, A.~Vogler, and C.~Weber.
\newblock Multi-day-ahead electricity price forecasting: A comparison of fundamental, econometric and hybrid models.
\newblock 2021.

\bibitem[de~Marcos et~al.(2019)de~Marcos, Bello, and Reneses]{marcos2019electricity}
R.~A. de~Marcos, A.~Bello, and J.~Reneses.
\newblock Electricity price forecasting in the short term hybridising fundamental and econometric modelling.
\newblock \emph{Electric Power Systems Research}, 167:\penalty0 240--251, 2019.

\bibitem[Pape et~al.(2016)Pape, Hagemann, and Weber]{pape2016fundamentals}
C.~Pape, S.~Hagemann, and C.~Weber.
\newblock Are fundamentals enough? explaining price variations in the german day-ahead and intraday power market.
\newblock \emph{Energy Economics}, 54:\penalty0 376--387, 2016.

\bibitem[Pfenninger and Pickering(2018)]{pfenninger2018calliope}
S.~Pfenninger and B.~Pickering.
\newblock Calliope: a multi-scale energy systems modelling framework.
\newblock \emph{Journal of Open Source Software}, 3\penalty0 (29):\penalty0 825, 2018.

\bibitem[Howells et~al.(2011)Howells, Rogner, Strachan, Heaps, Huntington, Kypreos, Hughes, Silveira, DeCarolis, Bazillian, et~al.]{howells2011osemosys}
M.~Howells, H.~Rogner, N.~Strachan, C.~Heaps, H.~Huntington, S.~Kypreos, et~al.
\newblock Osemosys: the open source energy modeling system: an introduction to its ethos, structure and development.
\newblock \emph{Energy Policy}, 39\penalty0 (10):\penalty0 5850--5870, 2011.

\bibitem[Schimeczek et~al.(2023)Schimeczek, Nienhaus, Frey, Sperber, Sarfarazi, Nitsch, Kochems, and El~Ghazi]{schimeczek2023amiris}
C.~Schimeczek, K.~Nienhaus, U.~Frey, E.~Sperber, S.~Sarfarazi, F.~Nitsch, et~al.
\newblock Amiris: Agent-based market model for the investigation of renewable and integrated energy systems.
\newblock \emph{Journal of Open Source Software}, 8\penalty0 (84):\penalty0 5041, 2023.

\bibitem[Harder et~al.(2025)Harder, Miskiw, Khanra, Maurer, Patil, Qussous, Weinhardt, Klobasa, Ragwitz, and Weidlich]{harder2025assume}
N.~Harder, K.~K. Miskiw, M.~Khanra, F.~Maurer, P.~Patil, R.~Qussous, et~al.
\newblock Assume: An agent-based simulation framework for exploring electricity market dynamics with reinforcement learning.
\newblock \emph{SoftwareX}, 30:\penalty0 102176, 2025.

\bibitem[Ghelasi and Ziel(2025)]{ghelasi2025day}
P.~Ghelasi and F.~Ziel.
\newblock From day-ahead to mid and long-term horizons with econometric electricity price forecasting models.
\newblock \emph{Renewable and Sustainable Energy Reviews}, 217:\penalty0 115684, 2025.

\bibitem[Bello et~al.(2016)Bello, Bunn, Reneses, and Mu{\~n}oz]{bello2016medium}
A.~Bello, D.~W. Bunn, J.~Reneses, and A.~Mu{\~n}oz.
\newblock Medium-term probabilistic forecasting of electricity prices: A hybrid approach.
\newblock \emph{IEEE Transactions on Power Systems}, 32\penalty0 (1):\penalty0 334--343, 2016.

\bibitem[Gonzalez et~al.(2011)Gonzalez, Contreras, and Bunn]{gonzalez2011forecasting}
V.~Gonzalez, J.~Contreras, and D.~W. Bunn.
\newblock Forecasting power prices using a hybrid fundamental-econometric model.
\newblock \emph{IEEE Transactions on Power Systems}, 27\penalty0 (1):\penalty0 363--372, 2011.

\bibitem[Gabrielli et~al.(2022)Gabrielli, W{\"u}thrich, Blume, and Sansavini]{gabrielli2022data}
P.~Gabrielli, M.~W{\"u}thrich, S.~Blume, and G.~Sansavini.
\newblock Data-driven modeling for long-term electricity price forecasting.
\newblock \emph{Energy}, 244:\penalty0 123107, 2022.

\bibitem[Liang and Dvorkin(2023)]{liang2023data}
Z.~Liang and Y.~Dvorkin.
\newblock Data-driven inverse optimization for marginal offer price recovery in electricity markets.
\newblock In \emph{Proceedings of the 14th ACM International Conference on Future Energy Systems}, pages 497--509, 2023.

\bibitem[Birge et~al.(2017)Birge, Horta{\c{c}}su, and Pavlin]{birge2017inverse}
J.~R. Birge, A.~Horta{\c{c}}su, and J.~M. Pavlin.
\newblock Inverse optimization for the recovery of market structure from market outcomes: An application to the miso electricity market.
\newblock \emph{Operations Research}, 65\penalty0 (4):\penalty0 837--855, 2017.

\bibitem[Chen et~al.(2017)Chen, Paschalidis, and Caramanis]{chen2017strategic}
R.~Chen, I.~C. Paschalidis, and M.~C. Caramanis.
\newblock Strategic equilibrium bidding for electricity suppliers in a day-ahead market using inverse optimization.
\newblock In \emph{2017 IEEE 56th Annual Conference on Decision and Control (CDC)}, pages 220--225. IEEE, 2017.

\bibitem[Ruiz et~al.(2013)Ruiz, Conejo, and Bertsimas]{ruiz2013revealing}
C.~Ruiz, A.~J. Conejo, and D.~J. Bertsimas.
\newblock Revealing rival marginal offer prices via inverse optimization.
\newblock \emph{IEEE Transactions on Power Systems}, 28\penalty0 (3):\penalty0 3056--3064, 2013.

\bibitem[Weron et~al.(2004)Weron, Bierbrauer, and Tr{\"u}ck]{weron2004modeling}
R.~Weron, M.~Bierbrauer, and S.~Tr{\"u}ck.
\newblock Modeling electricity prices: jump diffusion and regime switching.
\newblock \emph{Physica A: Statistical Mechanics and its Applications}, 336\penalty0 (1-2):\penalty0 39--48, 2004.

\bibitem[Haldrup and Nielsen(2006)]{haldrup2006regime}
N.~Haldrup and M.~O. Nielsen.
\newblock A regime switching long memory model for electricity prices.
\newblock \emph{Journal of econometrics}, 135\penalty0 (1-2):\penalty0 349--376, 2006.

\bibitem[Mount et~al.(2006)Mount, Ning, and Cai]{mount2006predicting}
T.~D. Mount, Y.~Ning, and X.~Cai.
\newblock Predicting price spikes in electricity markets using a regime-switching model with time-varying parameters.
\newblock \emph{Energy Economics}, 28\penalty0 (1):\penalty0 62--80, 2006.

\bibitem[Karakatsani and Bunn(2008)]{karakatsani2008intra}
N.~V. Karakatsani and D.~W. Bunn.
\newblock Intra-day and regime-switching dynamics in electricity price formation.
\newblock \emph{Energy Economics}, 30\penalty0 (4):\penalty0 1776--1797, 2008.

\bibitem[Kapoor et~al.(2023)Kapoor, Wichitaksorn, and Zhang]{kapoor2023analyzing}
G.~Kapoor, N.~Wichitaksorn, and W.~Zhang.
\newblock Analyzing and forecasting electricity price using regime-switching models: The case of new zealand market.
\newblock \emph{Journal of Forecasting}, 42\penalty0 (8):\penalty0 2011--2026, 2023.

\bibitem[Ziel and Weron(2018)]{ziel2018day}
F.~Ziel and R.~Weron.
\newblock Day-ahead electricity price forecasting with high-dimensional structures: Univariate vs. multivariate modeling frameworks.
\newblock \emph{Energy Economics}, 70:\penalty0 396--420, 2018.

\bibitem[Zhang et~al.(2020)Zhang, Tan, and Wei]{zhang2020adaptive}
J.~Zhang, Z.~Tan, and Y.~Wei.
\newblock An adaptive hybrid model for short term electricity price forecasting.
\newblock \emph{Applied Energy}, 258:\penalty0 114087, 2020.

\bibitem[Alkawaz et~al.(2022)Alkawaz, Abdellatif, Kanesan, Khairuddin, and Gheni]{alkawaz2022day}
A.~N. Alkawaz, A.~Abdellatif, J.~Kanesan, A.~S.~M. Khairuddin, and H.~M. Gheni.
\newblock Day-ahead electricity price forecasting based on hybrid regression model.
\newblock \emph{IEEE Access}, 10:\penalty0 108021--108033, 2022.

\bibitem[Yang et~al.(2019)Yang, Wang, Niu, and Du]{yang2019hybrid}
W.~Yang, J.~Wang, T.~Niu, and P.~Du.
\newblock A hybrid forecasting system based on a dual decomposition strategy and multi-objective optimization for electricity price forecasting.
\newblock \emph{Applied energy}, 235:\penalty0 1205--1225, 2019.

\bibitem[Shafie-Khah et~al.(2011)Shafie-Khah, Moghaddam, and Sheikh-El-Eslami]{shafie2011price}
M.~Shafie-Khah, M.~P. Moghaddam, and M.~Sheikh-El-Eslami.
\newblock Price forecasting of day-ahead electricity markets using a hybrid forecast method.
\newblock \emph{Energy Conversion and Management}, 52\penalty0 (5):\penalty0 2165--2169, 2011.

\bibitem[Wan et~al.(2013)Wan, Xu, Wang, Dong, and Wong]{wan2013hybrid}
C.~Wan, Z.~Xu, Y.~Wang, Z.~Y. Dong, and K.~P. Wong.
\newblock A hybrid approach for probabilistic forecasting of electricity price.
\newblock \emph{IEEE Transactions on Smart Grid}, 5\penalty0 (1):\penalty0 463--470, 2013.

\bibitem[Huang et~al.(2021)Huang, Shen, Chen, and Chen]{huang2021novel}
C.-J. Huang, Y.~Shen, Y.-H. Chen, and H.-C. Chen.
\newblock A novel hybrid deep neural network model for short-term electricity price forecasting.
\newblock \emph{International Journal of Energy Research}, 45\penalty0 (2):\penalty0 2511--2532, 2021.

\bibitem[Esser et~al.(2025)Esser, Finke, and Bertsch]{esser2025multi}
K.~Esser, J.~Finke, and V.~Bertsch.
\newblock A multi-objective inverse (moin) energy system modelling method for guiding early technology development.
\newblock \emph{Available at SSRN 5287540}, 2025.

\bibitem[Tarantola(2005)]{tarantola2005inverse}
A.~Tarantola.
\newblock \emph{Inverse problem theory and methods for model parameter estimation}.
\newblock SIAM, 2005.

\bibitem[Kirsch et~al.(2011)]{kirsch2011introduction}
A.~Kirsch et~al.
\newblock \emph{An introduction to the mathematical theory of inverse problems}, volume 120.
\newblock Springer, 2011.

\bibitem[da~Silva~Neto et~al.(2023)da~Silva~Neto, Becceneri, de~Campos~Velho, and Teixeira]{da2023computational}
A.~J. da~Silva~Neto, J.~C. Becceneri, H.~F. de~Campos~Velho, and R.~Teixeira.
\newblock \emph{Computational intelligence applied to inverse problems in radiative transfer}.
\newblock Springer, 2023.

\bibitem[Groetsch and Groetsch(1993)]{groetsch1993inverse}
C.~W. Groetsch and C.~Groetsch.
\newblock \emph{Inverse problems in the mathematical sciences}, volume~52.
\newblock Springer, 1993.

\bibitem[Richter(2015)]{richter2015inverse}
M.~Richter.
\newblock \emph{Inverse Probleme}.
\newblock Springer, 2015.

\bibitem[Gonz{\'a}lez and Bandera(2022)]{gonzalez2022building}
V.~G. Gonz{\'a}lez and C.~F. Bandera.
\newblock A building energy models calibration methodology based on inverse modelling approach.
\newblock In \emph{Building Simulation}, volume~15, pages 1883--1898. Springer, 2022.

\bibitem[Kurahashi(2018)]{kurahashi2018model}
S.~Kurahashi.
\newblock Model prediction and inverse simulation.
\newblock \emph{Innovative Approaches in Agent-Based Modelling and Business Intelligence}, pages 139--156, 2018.

\bibitem[Murray-Smith(2000)]{murray2000inverse}
D.~Murray-Smith.
\newblock The inverse simulation approach: a focused review of methods and applications.
\newblock \emph{Mathematics and computers in simulation}, 53\penalty0 (4-6):\penalty0 239--247, 2000.

\bibitem[Kov{\'a}cs(2021)]{kovacs2021inverse}
A.~Kov{\'a}cs.
\newblock Inverse optimization approach to the identification of electricity consumer models.
\newblock \emph{Central European Journal of Operations Research}, 29\penalty0 (2):\penalty0 521--537, 2021.

\bibitem[Esteban-Perez et~al.(2024)Esteban-Perez, Bunn, and Ghiassi-Farrokhfal]{esteban2024estimating}
A.~Esteban-Perez, D.~Bunn, and Y.~Ghiassi-Farrokhfal.
\newblock Estimating the unobservable components of electricity demand response with inverse optimization.
\newblock \emph{arXiv preprint arXiv:2410.02774}, 2024.

\bibitem[Nguyen(2020)]{nguyen2020cooperative}
D.~H. Nguyen.
\newblock Cooperative learning for p2p energy trading via inverse optimization and interval analysis.
\newblock \emph{arXiv preprint arXiv:2011.02609}, 2020.

\bibitem[Saez-Gallego et~al.(2016)Saez-Gallego, Morales, Zugno, and Madsen]{saez2016data}
J.~Saez-Gallego, J.~M. Morales, M.~Zugno, and H.~Madsen.
\newblock A data-driven bidding model for a cluster of price-responsive consumers of electricity.
\newblock \emph{IEEE Transactions on Power Systems}, 31\penalty0 (6):\penalty0 5001--5011, 2016.

\bibitem[Saez-Gallego and Morales(2017)]{saez2017short}
J.~Saez-Gallego and J.~M. Morales.
\newblock Short-term forecasting of price-responsive loads using inverse optimization.
\newblock \emph{IEEE Transactions on Smart Grid}, 9\penalty0 (5):\penalty0 4805--4814, 2017.

\bibitem[Gallego(2017)]{gallego2017inverse}
J.~S. Gallego.
\newblock Inverse optimization and forecasting techniques applied to decision-making in electricity markets.
\newblock 2017.

\bibitem[Bertsimas et~al.(2015)Bertsimas, Gupta, and Paschalidis]{bertsimas2015data}
D.~Bertsimas, V.~Gupta, and I.~C. Paschalidis.
\newblock Data-driven estimation in equilibrium using inverse optimization.
\newblock \emph{Mathematical Programming}, 153:\penalty0 595--633, 2015.

\bibitem[Fern{\'a}ndez-Blanco et~al.(2021)Fern{\'a}ndez-Blanco, Morales, Pineda, and Porras]{fernandez2021inverse}
R.~Fern{\'a}ndez-Blanco, J.~M. Morales, S.~Pineda, and {\'A}.~Porras.
\newblock Inverse optimization with kernel regression: Application to the power forecasting and bidding of a fleet of electric vehicles.
\newblock \emph{Computers \& Operations Research}, 134:\penalty0 105405, 2021.

\bibitem[Chen et~al.(2019)Chen, Paschalidis, Caramanis, and Andrianesis]{chen2019learning}
R.~Chen, I.~C. Paschalidis, M.~C. Caramanis, and P.~Andrianesis.
\newblock Learning from past bids to participate strategically in day-ahead electricity markets.
\newblock \emph{IEEE Transactions on Smart Grid}, 10\penalty0 (5):\penalty0 5794--5806, 2019.

\bibitem[Coester et~al.(2018)Coester, Hofkes, and Papyrakis]{coester2018optimal}
A.~Coester, M.~W. Hofkes, and E.~Papyrakis.
\newblock An optimal mix of conventional power systems in the presence of renewable energy: A new design for the german electricity market.
\newblock \emph{Energy Policy}, 116:\penalty0 312--322, 2018.

\bibitem[Beran et~al.(2019)Beran, Pape, and Weber]{beran2019modelling}
P.~Beran, C.~Pape, and C.~Weber.
\newblock Modelling german electricity wholesale spot prices with a parsimonious fundamental model--validation \& application.
\newblock \emph{Utilities Policy}, 58:\penalty0 27--39, 2019.

\bibitem[Bischl et~al.(2017)Bischl, Richter, Bossek, Horn, Thomas, and Lang]{mlrMBO}
B.~Bischl, J.~Richter, J.~Bossek, D.~Horn, J.~Thomas, and M.~Lang.
\newblock \emph{{{{mlrMBO}}: {{A Modular Framework}} for {{Model}}-{{Based Optimization}} of {{Expensive Black}}-{{Box Functions}}}}, 2017.
\newblock URL \url{https://arxiv.org/abs/1703.03373}.

\bibitem[Eddelbuettel et~al.(2025)Eddelbuettel, Francois, Allaire, Ushey, Kou, Russell, Ucar, Bates, and Chambers]{rcpp}
D.~Eddelbuettel, R.~Francois, J.~Allaire, K.~Ushey, Q.~Kou, N.~Russell, et~al.
\newblock \emph{Rcpp: Seamless R and C++ Integration}, 2025.
\newblock URL \url{https://CRAN.R-project.org/package=Rcpp}.
\newblock R package version 1.1.0.

\bibitem[Mersmann(2024)]{microbenchmark}
O.~Mersmann.
\newblock \emph{microbenchmark: Accurate Timing Functions}, 2024.
\newblock URL \url{https://CRAN.R-project.org/package=microbenchmark}.
\newblock R package version 1.5.0.

\end{thebibliography}

\end{document}